\documentclass[twocolumn,superscriptaddress,amssymb, nobibnotes, aps, prb]{revtex4-2}
\usepackage{enumitem}
\usepackage{amsthm}
\usepackage{hhline}
\usepackage{epsfig,amsopn}
\usepackage{graphicx}
\usepackage{xcolor}
\usepackage{amsmath,amssymb}
\usepackage{url}
\usepackage{float}
\usepackage{enumerate}
\usepackage{verbatim}
\usepackage{bm}

\newcommand{\fref}[1]{Fig.~\ref{#1}}

\newcommand{\eref}[1]{Eq.~(\ref{#1})}

\newcommand{\cref}[1]{chapter~\ref{#1}}

\newcommand{\Cref}[1]{Chapter~\ref{#1}}

\newcommand{\aref}[1]{Appendix~\ref{#1}}

\begin{document}
\title{Transmission in a Fano-Anderson chain with a topological defect} 

\author{Ritu Nehra}
\affiliation{Department of Physics, Indian Institute of Science Education and Research, Bhopal, Madhya Pradesh 462066, India}
\affiliation{Raman Research Institute, Bangalore, Karnataka 560080, India}
\author{Ajith Ramachandran}
\author{Sebastian W\"{u}ster}
\author{Auditya Sharma}
\affiliation{Department of Physics, Indian Institute of Science Education and Research, Bhopal, Madhya Pradesh 462066, India}

\begin{abstract}
The Fano-Anderson chain consists of a linear lattice with a discrete
side-unit, and exhibits Fano-resonant scattering due to coupling
between the discrete states of the side-unit with the tight-binding
continuum. We study Fano-resonance-assisted transport for the case of
a topologically non-trivial side unit. We find that the topology of
the side unit influences the transmission characteristics which thus
can be an effective detection tool of the topological phases of the
side unit. Furthermore, we explore the role of dual links between the
linear tight binding chain and the side unit. The secondary connection
between the main chain and the side unit can modify the position or
width of the Fano resonance dip in the transmission probability, and
thus yield additional control.
\end{abstract}

\maketitle

\section{Introduction}
Fano resonance~\cite{fano1935sullo,fano1961effects} is a quantum
phenomenon arising from the interaction between a continuum band of
states and discrete states, that has been explored in a wide variety
of settings ranging from nuclear, atomic, molecular and condensed
matter systems \cite{miroshnichenko2010fano,ujsaghy2000theory}, to
electronics~\cite{rotter2004tunable,attaran2014circuit} and optics
\cite{fan2014optical,argyropoulos2013plasmonic,shafiei2013subwavelength}. Linear
wave equations on Hamiltonian lattices can also give rise to Fano
resonances if local defects are present
\cite{miroshnichenko2010fano,ramachandran2018fano}. The Fano-Anderson
chain made up of a linear chain interacting with a single defect site
is the simplest example for such a system
\cite{miroshnichenko2005engineering}. The discrete state introduced by
the defect allows additional propagation paths for scattering waves
which interfere constructively or destructively. This
discrete-state-assisted interference may lead to perfect transmission
or perfect reflection along with an asymmetric resonance profile.

Recent years have seen quickly growing interest in the exploration of
Fano resonance associated with many different kinds of modified
Fano-Anderson chains. Examples include systems with a single defect
site \cite{longhi2007bound,bulgakov2009resonance}, a linear chain
\cite{chakrabarti2007fano}, and a Fibonacci chain
\cite{chakrabarti2006electronic}. One common observation in all these
structures is that the Fano resonance profile in the transmission
probability shows a dip at the eigen-energies of the side unit
interacting with the main chain. Since topological
systems~\cite{lu2014topological,st2017lasing} are naturally endowed
with isolated energy states, one expects interesting Fano scattering properties. Nonetheless, Fano resonances in such
topological models have only very recently drawn the first attention~\cite{zangeneh2019topological}.

We take the Su-Schieffer-Heeger (SSH) chain as an example for the
topological side unit and demonstrate how the Fano transmission
profile is affected by the topological-to-trivial phase transition.
In the trivial phase, a perfect transmission is observed at energies
close to $E = 0$.  The isolated edge state in the topological phase
induces a Fano resonance and leads to a dip in the transmission
profile at the energy of the edge state. The trivial to topological
transition can be identified from the dip emerging in the transmission
profile. A second connection to the side unit enhances the width of
the resonance dip, which may be an advantage in real detection
systems. The boundary conditions of the SSH chain modify the edge
states and the transmission profile is also altered accordingly.  This
concept can be further extended to two or three dimensional
topological systems.

In this article, we employ the transfer matrix method (TMM) to study
the transmission profile~\citep{PhysRevB.59.8639,miroshnichenko2005engineering} of the Fano-Anderson chain, which allows for
exact analytical expressions describing the transmission
characteristics. The method has been used in the past, to explore the
Fano-Anderson chain with one connection to the side
unit~\cite{miroshnichenko2010fano}. We focus on a linear chain with
two connections to a topological side unit. We derive the exact
expression for the transmission coefficient and obtain the conditions
for perfect reflection. Our methods are general enough to be
applicable to any kind of finite sized side unit, which may be quasi
one-dimensional, two-dimensional, or even three-dimensional.

\begin{figure}[h!]
\centering
\hspace{-14pt}\includegraphics[width=1.05\linewidth,height=0.425\linewidth]{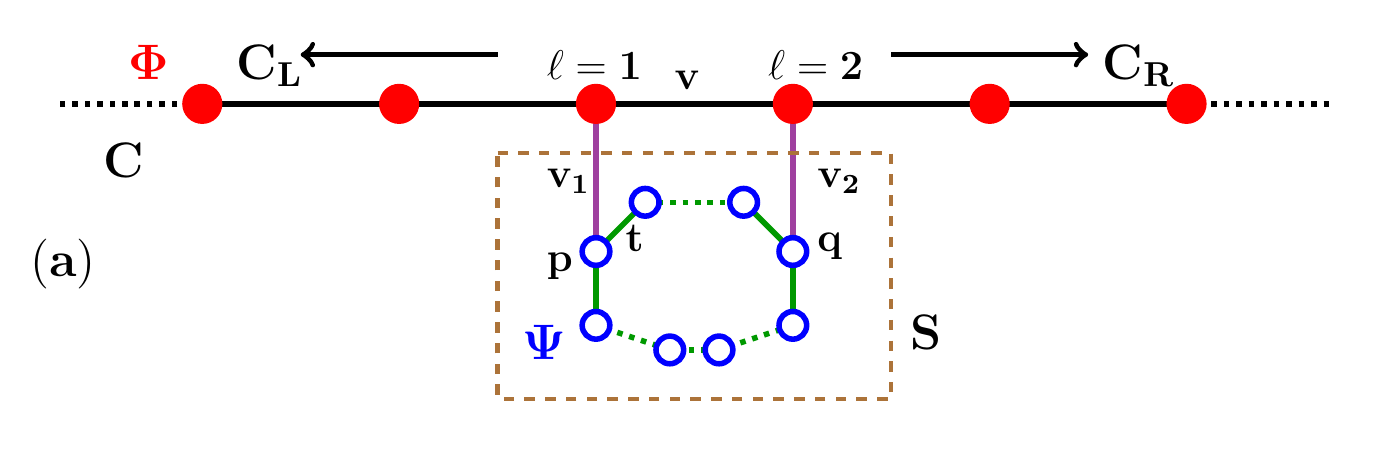}
\hspace{-14pt}\includegraphics[width=1\linewidth,height=0.175\linewidth]{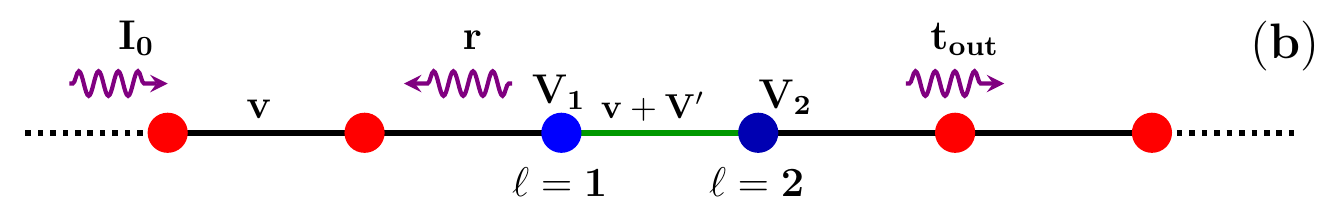}
\caption{(a) Schematic of the setup with two adjacent connections between the
  main chain (C) and the $p^{th}$ and $q^{th}$ sites of the side unit (S). These connections have hopping strengths $v_1$ and $v_2$. The hopping amplitude in the
  main chain is set to $v$ with onsite wavefunction $\Phi$ and hopping in the side unit $t$. We mark the left and
  right parts of the main chain as $C_L$ and $C_R$ and side-unit as $S$ for clarity. (b) Schematic representation of
  Fano-Anderson chain with effective potentials $V_1$ and $V_2$ at sites $\ell=1$ and $\ell=2$, respectively, which result after elimination of the side-unit. The effective hopping between these two sites becomes $v+V'$.}
\label{structure}
\end{figure}

The manuscript is arranged as follows. In the next section we describe
the Fano-Anderson model which is linked to a side unit with two
adjacent connections. This section also describes how the transfer
matrix method can be used to obtain the conditions needed for Fano
resonance. In section~\ref{sec3}, we present the topological
properties of the SSH model considering all possible
structures. Further, in section~\ref{sec4}, we study Fano resonance in
different cases of the SSH model taken as the side unit.  Finally, we
collect and summarize our findings in the conclusion section.


\section{\label{sec2}Model}
We consider a linear tight binding chain connected to a Fano defect
chain with two connections between them as shown in
\fref{structure}. The Hamiltonian for this system is 
\begin{equation}
H = H_c + H_s + H_{cs},
\label{ham}
\end{equation}
where $H_c$ is the Hamiltonian corresponding to the linear chain $C$, 
\begin{equation}
H_c = v  \sum_{\ell}(\Phi^*_{\ell + 1}\Phi_{\ell} + \Phi^*_{\ell}\Phi_{\ell + 1}).
\label{ham_c}
\end{equation}
Here, $\Phi_{\ell}$ is the complex wave amplitude at the site $\ell$
and $v$ is the coupling strength. $H_s$ corresponds to the Hamiltonian
for the side chain $S$ with $N$ sites and reads
\begin{equation}
H_s =  \sum_{i , j = 1}^{N}h_{i , j}\Psi^*_i\Psi_j,
\label{ham_s}
\end{equation}
where $\Psi_i$ is the complex wave amplitude at the site $i$ on the
side chain and $h_{ij}$ is the coupling strength connecting the sites
labeled $i, j$. The third term in the Hamiltonian $H_{cs}$
corresponds to the coupling between the main chain and the side chain,
and is given by:
\begin{equation}
H_{cs} = v_1 \Phi_{1}^{*}\Psi_{p} + v_2 \Phi_{2}^{*}\Psi_{q} + H.c.,
\label{ham_cs}
\end{equation}
with $v_1$ and $v_2$ being the coupling strengths between the main
chain and the side chain. Sites $p$ and $q$ are arbitrary and even can
be the same.  In the absence of the side unit, Bloch's theorem can be
applied to the translationally invariant linear chain $C$ described by
\eref{ham_c}. For states $\phi_\ell = e^{\iota k\ell}$ one finds
energies of propagating waves $E=2v \cos{k}$, where $k$ is the wave
number. The side unit acts as a scatterer and introduces extra paths
for an incoming wave. This resonant scattering due to the side unit
controls the wave propagation in the main chain and we employ the
transfer matrix method to obtain the transmission and reflection
coefficients associated with the propagation in the main chain $C$ in
the presence of a side unit $S$.

The time ($\tau$) dependence from the Schr\"odinger equation is eliminated using the Ansatz 
\begin{align}
\label{t_d1}\Phi(\ell,\tau)=&A_{\ell}e^{-\iota E\tau},\\
\label{t_d2}\Psi(j,\tau)=&B_je^{-\iota E\tau}.
\end{align}
The time-independent Schr\"odinger equation corresponding to the Hamiltonian (H) can be written as:
\begin{align}
EA_{\ell} &= v(A_{\ell+1}+A_{\ell-1})+ v_1B_{p}\delta_{\ell,1} + v_2B_{q}\delta_{\ell,2}, \label{latt_eqc} \\
EB_{i} &= \sum_{j=1}^{N}h_{i,j}B_{j}+v_1\delta_{i,p}A_1+v_2\delta_{i,q}A_2 \label{latt_eqs},
\end{align}
where \eref{latt_eqc} and \eref{latt_eqs} correspond to sites in the main chain and side unit, respectively.  To obtain the
transmission characteristics in the main chain, the two coupled
equations must be solved simultaneously.  The complex amplitudes $B_i$
in the side unit can be written in terms of the amplitudes $A_\ell$ in
the main chain as follows. Using \eref{latt_eqs}, the relation between
$B_i$ and $A_\ell$ can be written as $\mathcal{H}_N \mathcal{B}_N =
\mathcal{A}_N$ where $\mathcal{H}_N(E) = [E I - H_s]_{N\times N}$,
$\mathcal{B}_N = [B_1,\hdots,B_p,\hdots,B_q,\hdots,B_N]^T$,
$\mathcal{A}_N=[0,\hdots,v_1A_1,\hdots,v_2A_2,\hdots,0]^T$, with $I$
the identity matrix and $N$ the size of the side chain (S). Inverting
the matrix $\mathcal{H}_N$~\cite{westlake1968handbook}, the
expressions for $B_{p/q}$ now become
\begin{align}
\label{psin}B_p=&v_1\alpha_{pp}A_1+v_2\alpha_{pq}A_2,\\
\label{psim}B_q=&v_1\alpha_{qp}A_1+v_2\alpha_{qq}A_2,
\end{align}
where $\alpha_{ij}(E)=({\mathcal{H}}^{-1}_N(E))_{ij}$. The Hermiticity
of the Hamiltonian guarantees that $\alpha_{ij} =
\alpha^*_{ji}$. These equations can be substituted back into
 \eref{latt_eqc} to obtain the wave propagation in the main chain in
terms of the amplitudes $A_l$ as:
\begin{align}
\hspace{-2pt}EA_\ell\hspace{-2pt}=\hspace{-2pt}(\alpha_{pp}v_1^2\delta_{\ell,1}\hspace{-2pt}+\hspace{-2pt}\alpha_{qq}v_2^2\delta_{\ell,2})A_{\ell}\hspace{-2pt}+\hspace{-2pt}(v\hspace{-2pt}+\hspace{-2pt}v_1v_2\alpha_{pq}\delta_{\ell,1})A_{\ell+1}\nonumber\\+(v\hspace{-2pt}+\hspace{-2pt}v_2v_1\alpha_{qp}\delta_{\ell,2})A_{\ell-1}.
\label{latt_eqnfinal}
\end{align}
Equation (\ref{latt_eqnfinal}) is the modified lattice equation for
the main chain incorporating the effects of the scatterer in terms of
effective energy dependent potentials $V_1(E) = \alpha_{pp}v_1^2$,
$V_2(E)=\alpha_{qq}v_2^2$, and $V'(E) = v_1v_2\alpha_{pq}$ as sketched
in \fref{structure}b. The energy dependence of the effective
potentials allows for resonant scattering and can cause complete
transmission or complete reflection of the incoming wave as well as an
asymmetric resonance profile. From \fref{structure}b, the condition
for perfect reflection can be identified as $v = -V'(E)$ since the
effective hopping between sites $\ell = 1$ and $\ell = 2$ would then
vanish. In the limit $v_2\rightarrow 0$ i.e. with just a single
connection between the main chain and the side unit, $V_2(E)$ and
$V'(E)$ vanish. Perfect transmission in this case occurs when
$|\mathcal{H}_{N-1}|=0\;\text{or}\;V_1=\alpha_{pp}=0$ which results in
a translationally invariant tight binding chain with no effective
scattering and therefore unit transmission. Perfect reflection is
instead realized when
$|\mathcal{H}_N|=0\;\text{or}\;V_1=\alpha_{pp}=\infty$ i.e. an
infinite onsite potential leads to full reflection or zero
transmission~\cite{miroshnichenko2005engineering}.

Employing the transfer matrix method (\aref{TMM}) for \eref{latt_eqnfinal}, the transmission coefficient in terms of the energy of the incoming wave can be written as: 
\begin{equation}
T=\frac{v^2(4v^2-E^2)|v+V'|^2}{[\beta v^2+\gamma v E-E^2 M^{'}_{11} M^{'}_{22}]},
\label{trans_2}\end{equation}
where $\beta=(M^{'}_{11}+M^{'}_{22})^2+(M^{'}_{12}-M^{'}_{21})^2$ and $\gamma=(M^{'}_{11}-M^{'}_{22})(M^{'}_{12}-M^{'}_{21})$ with $M^{'}_{11}=[(E-V_{1})(E-V_{2})-|v+V'|^2]$, $M^{'}_{12}=-v(E-V_{2})$, $M^{'}_{21}=v(E-V_{1})$, $M^{'}_{22}=-v^2$.
When two side chain sites are simultaneously connected to the main
chain, perfect reflection is possible if the energy is tuned in such 
a way that the condition
\begin{equation}
v+V'=0
\label{ref_eq}
\end{equation} 
holds, as can directly be seen in \eref{trans_2}, besides the
intuitive way of obtaining it from  \fref{structure}(b).  The condition given in
\eref{ref_eq} is a consequence of destructive interference of the
different paths available for the wave which results in zero hopping
amplitude for transmission across the scattering zone, leading to
perfect reflection of the wave.


\section{\label{sec3}SSH chain and topology}
\begin{figure}[h!]
\includegraphics[scale=0.85]{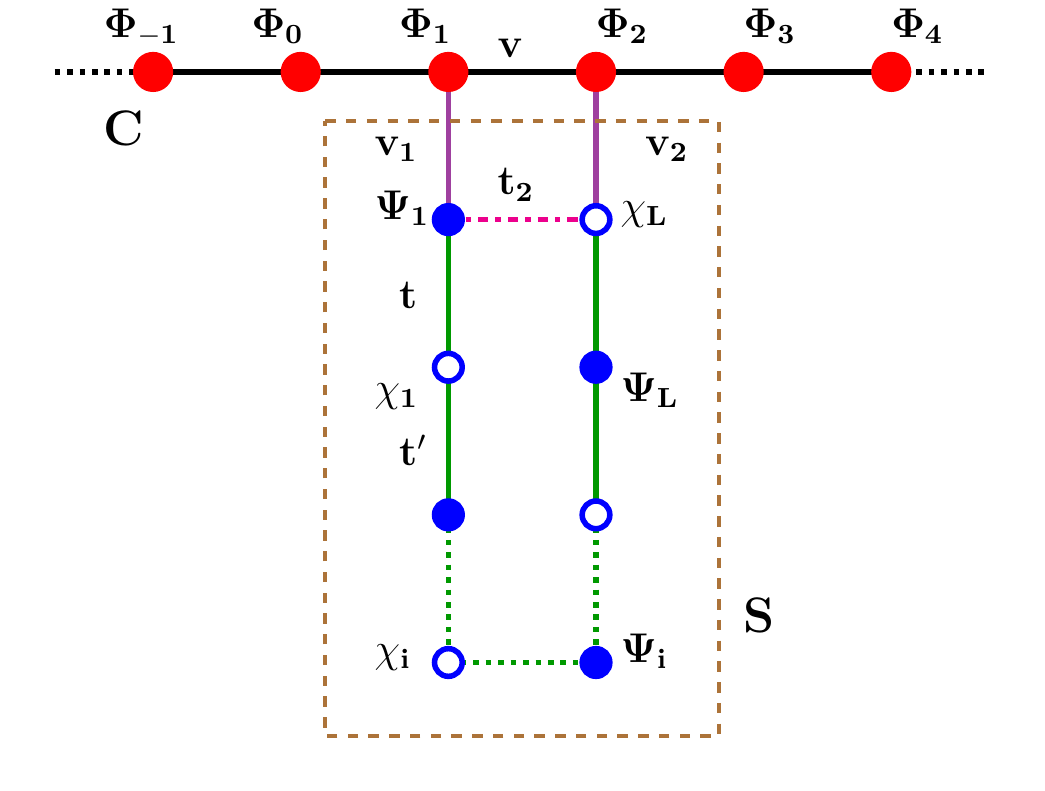}
\caption{Representation of the site amplitudes of wavefunctions, with
  the side unit described by a SSH model with inter-cell hopping $t'$ and
  intra-cell hopping $t$. The main chain (C) hopping, and the two
  connection hoppings are set to $v$, $v_1$, and $v_2$
  respectively. The presence (absence) of the extra hopping $t_2=t'$ with
  dashdotted magenta line turns the side unit into a closed (open)
  chain.}
\label{FA_ssh}
\end{figure}

In the discussion so far, details of the side unit did not enter. Now
we choose the Su-Schieffer-Heeger (SSH) model as the side unit because
of its interesting topological properties. The SSH model consists of a
network of sites with alternating hopping amplitudes with the
Hamiltonian given by
\begin{align}
H_s^{SSH}=\displaystyle\sum_{i=1}^{L}(t \Psi^{*}_{i}\chi_{i}+t'\chi^{*}_{i}\Psi_{i+1})+H.c.
\label{eq:ssh}
\end{align} 
where, $\Psi_i$, $\chi_i$ are the particle wavefunctions on two adjacent
sites of the $i^{th}$ unit cell. For the case of our side unit this is sketched in the dotted box of \fref{FA_ssh}. 
The intra-cell coupling is labeled as $t$ whereas inter-cell coupling is $t'$.  
\begin{figure*}
\centering
\includegraphics[scale=0.465]{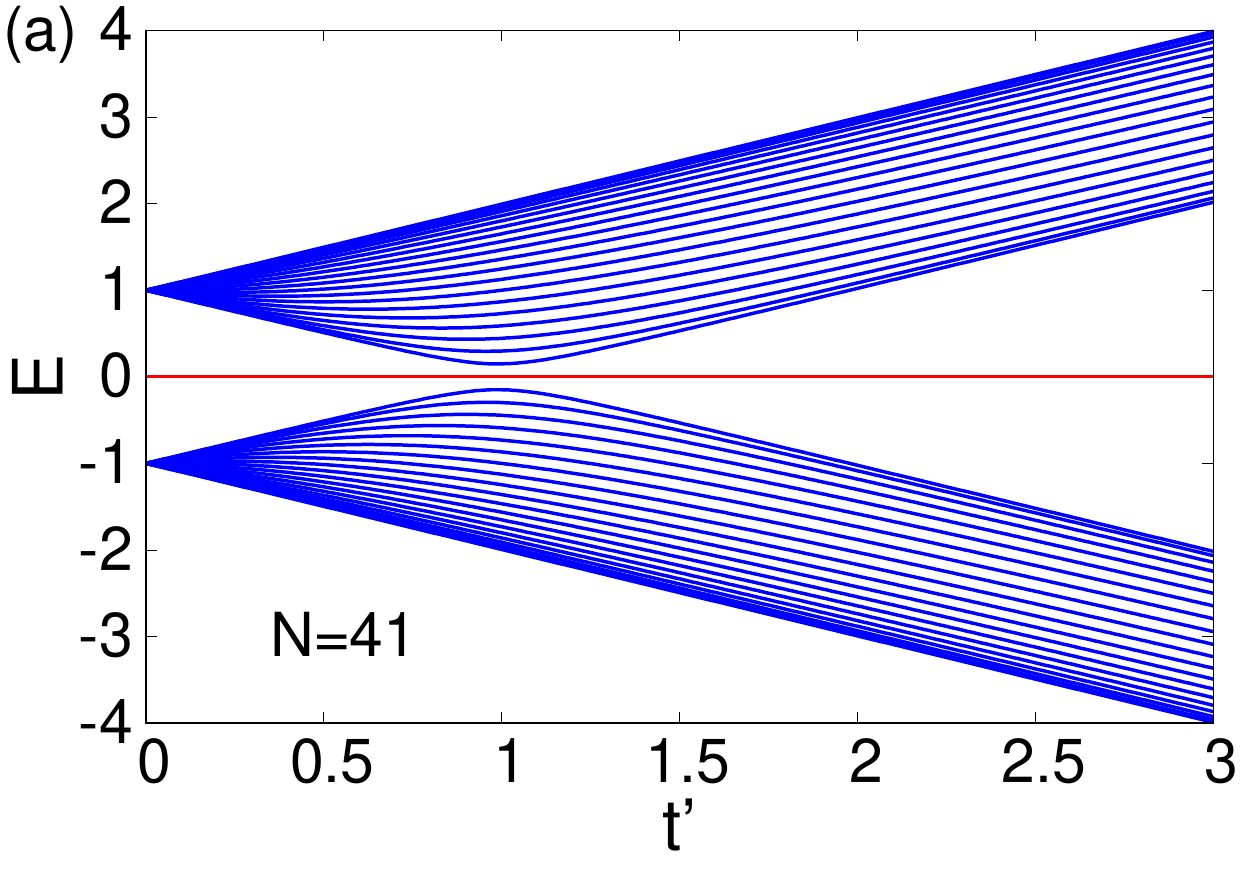}
\includegraphics[scale=0.465]{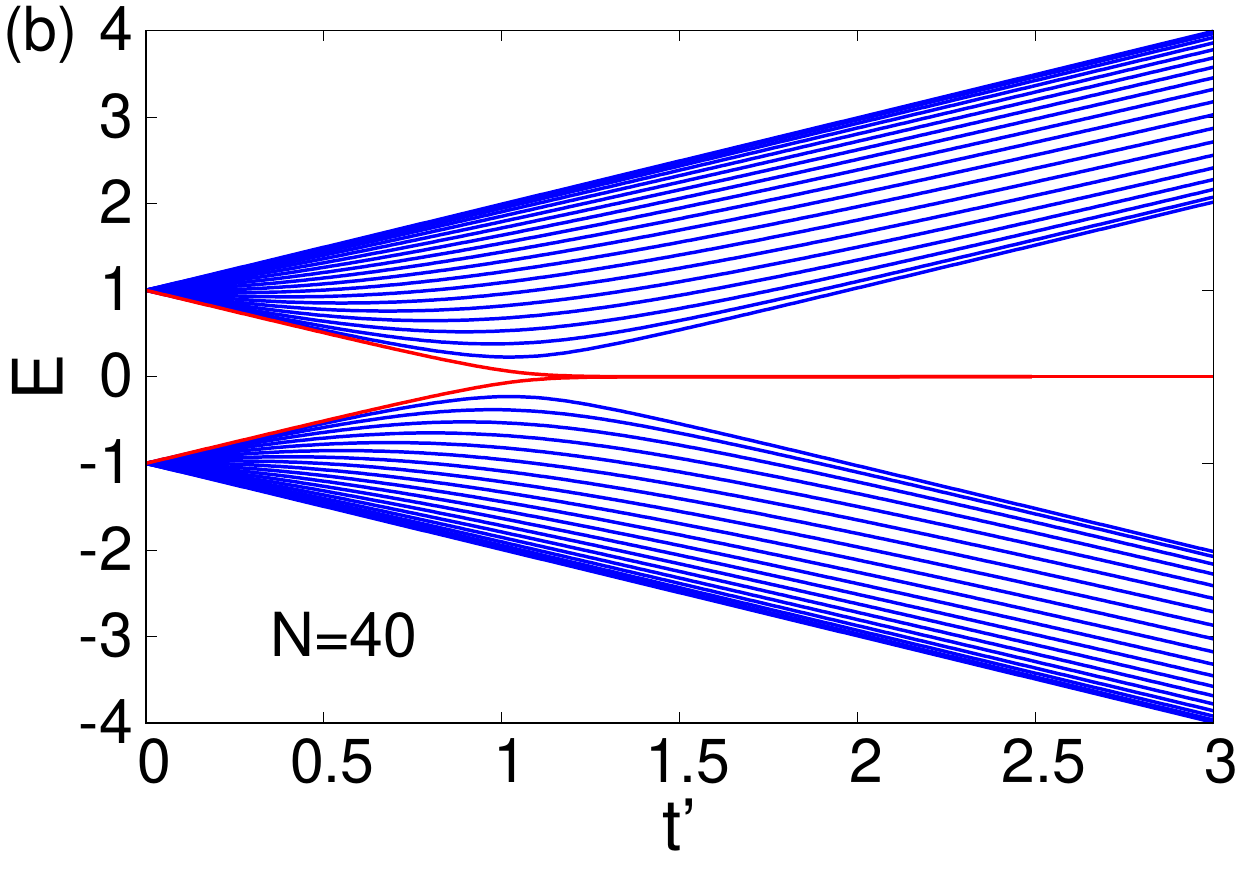}
\includegraphics[scale=0.465]{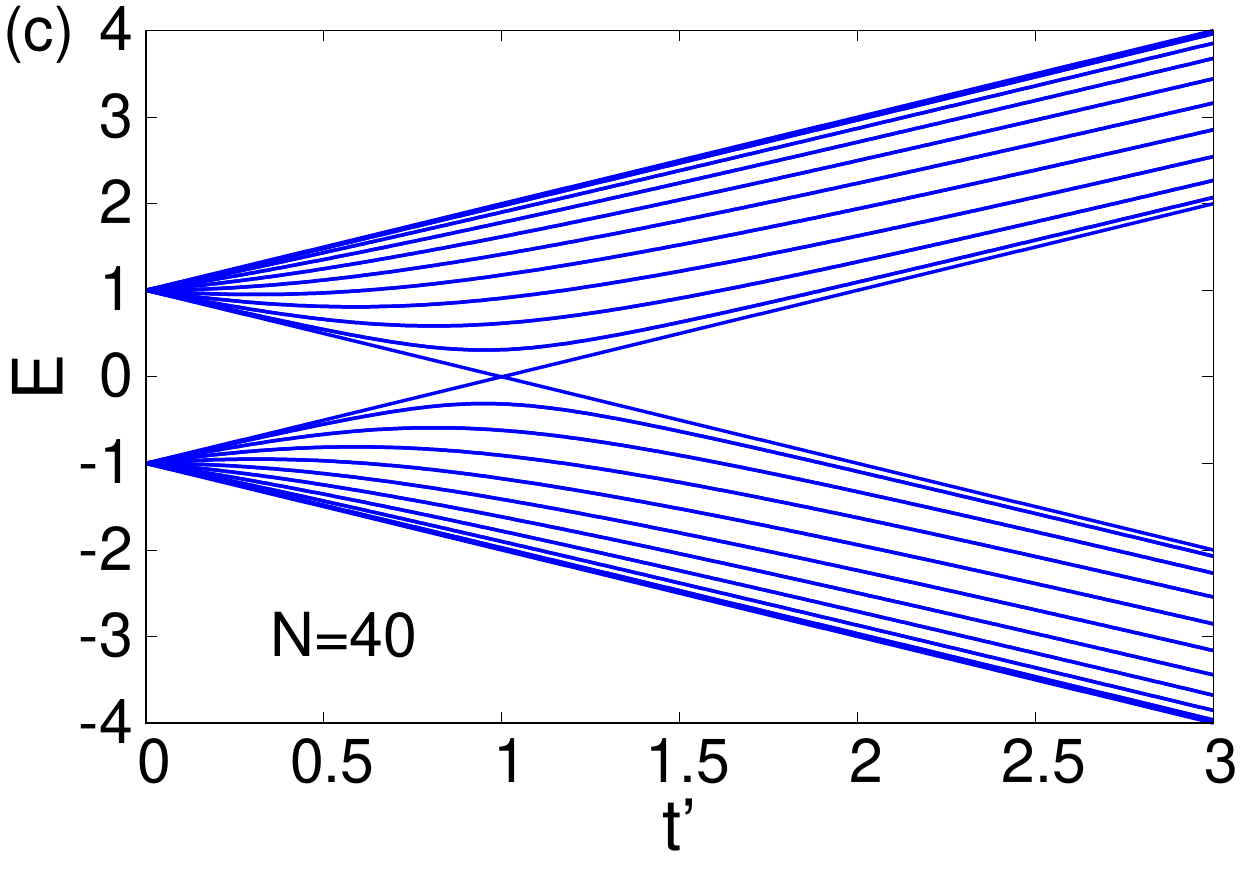}
\caption{The energy spectrum for the isolated SSH model with different
  inter-cell hopping $t'$ and a constant intra-cell hopping
  $t=1.0$ with red lines showing the energy of edge states. We show the spectrum for (a) $N=2L\pm1=41$ sites and (b)
  $N=2L=40$ sites with open boundary conditions and (c) $N=2L=40$ sites 
  with periodic boundary conditions.}
\label{en_ssh} 
\end{figure*}

The SSH model is known for its topological features. The winding
number changes from zero to one as the system is tuned from $t'<t$ to
$t'>t$, indicating a phase transition from the topological 
phase to a trivial phase. The dispersion spectrum of the SSH model for
different total number of sites $N=2L\pm1\;\text{or}\;2L$ is depicted
in \fref{en_ssh}. A chain with an odd number of sites is incompatible
with periodic boundary conditions. On the other hand an open chain may
have an odd or even number of sites and always gives rise to an edge
state with energy $E\approx0$ in the topological phase. The number of
zero energy edge states residing on the ends of the chain depends on
the way the chain is terminated at the boundaries.

For odd number of sites ( \fref{en_ssh}a), there is always an edge
state in the system. When $t'<t$ the edge state lies at the end with
the hopping $t'$ whereas for $t'>t$ the edge state lies at the end
with the hopping $t$ of the SSH chain. The ratio of the intra-cell to
the inter-cell hopping ($\frac{t}{t'}$) of the SSH chain controls the
position of this edge state. In contrast, for even number of sites,
edge states at both ends appear only when $t'>t$.  The edge states
present in the SSH model reside on the ends of the chain and decay in
the bulk depending on the ratio of the inter-cell and intra-cell
hopping ($\xi \propto\frac{t'+t}{t'-t}$). The system size should be
greater than the decay length for the properties of the edge states to
be observable.


\section{\label{sec4}Fano resonance: SSH chain as the side unit}
After a review of the physics of the isolated SSH model, we now
explore how coupling it to the main chain affects transport properties on
that chain. Here, we take the side unit to be the SSH chain as shown
in \fref{FA_ssh}. The expression for $\Gamma_N=|\mathcal{H}_N(E)|=|E
\mathcal{I}-H_s|$ is thus given by
\begin{equation}
\Gamma_N=|E \mathcal{I}-H_s|=
\begin{vmatrix}
E &-t  &0 &0 &\hdots &-t_2\\
-t &E &-t'&0 &\hdots &0\\
0 &-t' &E &-t  &\hdots &\vdots\\ 
0 &0 &\ddots &\ddots &\ddots &-t_1\\
-t_2 &0 &0 &\hdots &-t_1&E
\end{vmatrix}_{N\times N} 
\label{rho_ssh}
\end{equation}
where $N=2L$ or $2L\pm1$ is the number of sites in the SSH side unit
(with $L$ unit cells on the SSH chain) and $t$, $t'$ are two alternate
hoppings. Since $t$ and $t'$ alternate on the sub-diagonal and
super-diagonal of the tridiagonal matrix ($\mathcal{H}_N(E)$), one has to write  $t_1=t$
or $t'$ depending on the length $N$, while $t_2=0$ for an open chain
and $t_2=t'$ for a closed chain.  While the length of the open chain
may be odd or even, the closed chain is forced to have an even number
of sites, because of geometrical constraints, as we will discuss
later. We use
$\Gamma_{N-j}^{\{\lambda_1,\lambda_2,\hdots,\lambda_j\}}$ to represent
the determinant of the square matrix with dimension $(N-j) \times
(N-j)$ that is formed after the removal of the
$\{\lambda_1,\lambda_2,\hdots,\lambda_j\}^{th}$ rows and columns from
the matrix $[E\mathcal{I}-H_s]$ given in \eref{rho_ssh}.

 In the following subsections, we separately discuss how the system
 size of the side-unit and whether the SSH chain is open or closed
 affects transport in the main chain. The properties of the open and
 closed SSH chains are quite different due to the presence and
 absence, respectively, of edge states. The focus of this study is on
 understanding how the transmission properties carry signatures of
 these edge-states, which play a crucial role at energies close to
 zero ($E\sim 0$). Therefore, the first and last site ($n^{th}$) of
 the SSH chain are connected to the main chain as shown in
  \fref{FA_ssh}. 


\subsection{Case 1: Open chain with  odd N}
We now discuss the transport properties of the system when the defect
is the open SSH chain with an odd number of sites (\fref{FA_ssh}). As
discussed already, the open chain always possesses one edge state
whose location depends on the relative hopping strengths $t,\;t'$ as
shown in \fref{en_ssh}a. When $t'<t$, an
edge state resides on the last site of the SSH chain and decays in the
bulk while in the other limit $t' > t$, the edge state resides on the
first site. Besides varying the relative strength of $t$ and $t'$, we
have also considered two different cases for the second connection
strength $v_2$ i.e. $v_2=0$ and $v_2\neq 0$. For all scenarios we need
to first evaluate the required coefficients of the inverse of the
matrix $\mathcal{H}_N$. For the particular choice here, we have the
relations $\alpha_{11}=\frac{\Gamma_{N-1}^{\{1\}}}{\Gamma_{N}}$,
$\alpha_{nn}=\frac{\Gamma_{N-1}^{\{N\}}}{\Gamma_{N}}$,
$\alpha_{1n}=\alpha_{n1}=\frac{(tt')^{\frac{N-1}{2}}}{\Gamma_{N}}$.
\begin{figure}[h!]
\includegraphics[scale=1.15]{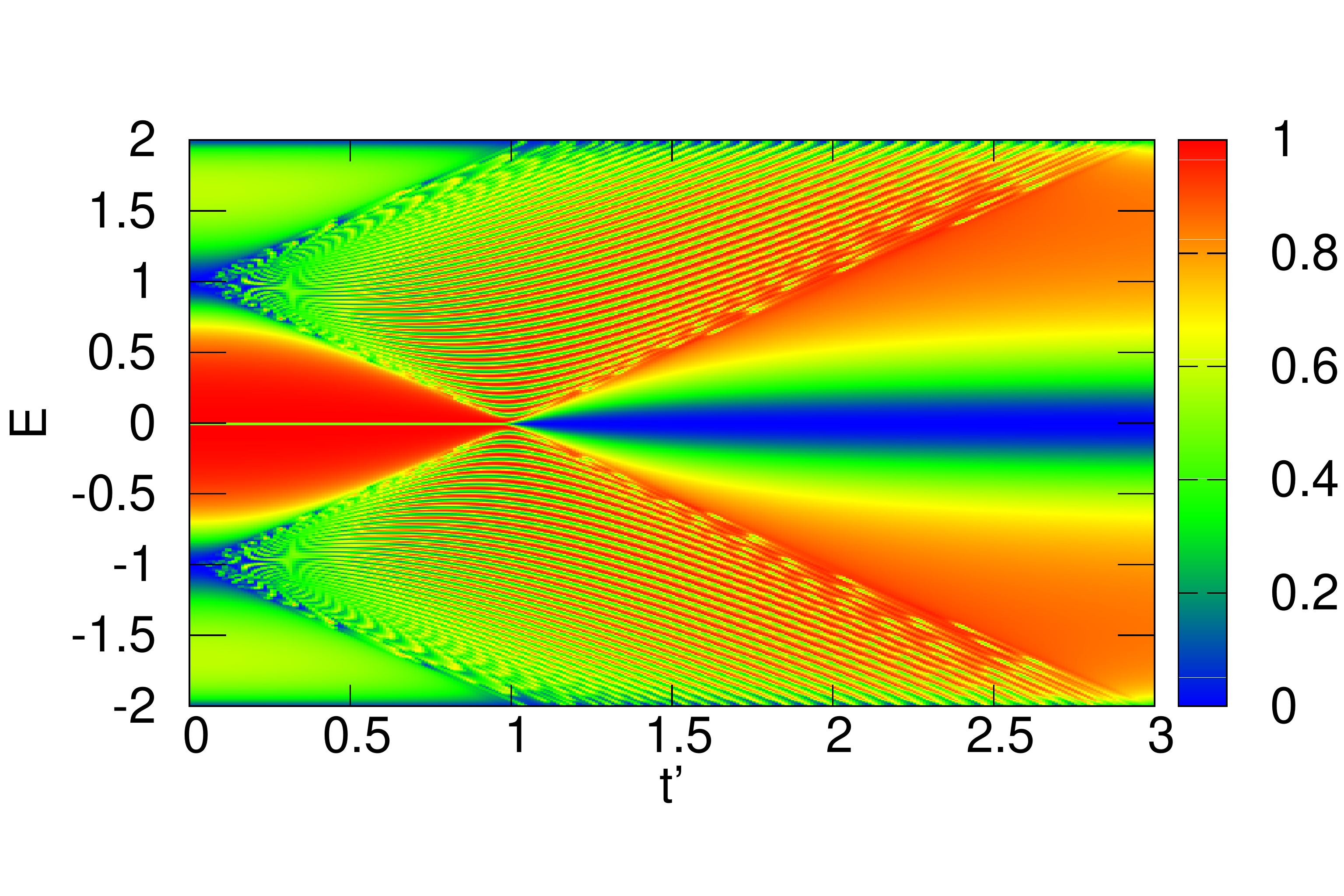}
\caption{The transmission coefficient $T$ for the open SSH side chain
  with $N=99$ for a single connection between chain and side unit. We show $T$ as a function of inter-cell hopping $t'$ and incoming wave energy $E$. The other parameters are kept constant:
  $v=1$, $t=1$, $v_1=1$ and $v_2=0$.}
\label{Ts_oo}
\end{figure}

First, we discuss the effect of a single link between the main chain
and the side unit i.e. $v_1\neq 0;\;v_2=0$. For this scenario, perfect
reflection ($T=0$) results if $\Gamma_N=0$ while perfect transmission
($T=1$) is found for $\Gamma_{N-1}^{\{1\}}=0$. Figure \ref{Ts_oo}
features the transmission coefficient (T) as a function of the
incoming energy of the wavefunction (E) and inter-cell hopping $t'$.
 It is observed that for $E=0$, the system shows a clear perfect
 reflection for $t'> t$ with a broad blue region. However, for $t'<t$
 a thin line is observed within the red region of full transmission as
 depicted in  \fref{Ts_oo}. The behavior of the transmission
coefficient (T) in this case is consistent with the
characteristics of the eigenstate of the full Hamiltonian at $E=0$ as
shown in \fref{wfxn_tbo_1} of \aref{wavefn}. In the limit $t'<t$,
  the eigenstate at $E=0$ supports no probability amplitude in the
  main chain and as a consequence, a sharp dip indicating zero transmission
   is observed as shown in
  \fref{Ts_oo}. In the limit $t'>t$, for the $E=0$ state, the
  probability amplitude resides only on one half of the main chain and
  hence, leads to perfect reflection as indicated by the broad blue
  region in \fref{Ts_oo}. 

If the second coupling to the Fano defect side chain is added
($v_2\neq 0$), the condition for perfect reflection is modified. It is
now possible only for those energies that satisfy:
\begin{equation}
v\Gamma_N+v_1v_2(tt')^{\frac{N-1}{2}}=0,
\label{eq22}
\end{equation}
where $\Gamma_{N}=\displaystyle\prod_{\theta_m}(E\pm\sqrt{t^2+{t'}^2+2tt'\cos\theta_m})$ with $\theta_m=\frac{2\pi m}{N+1}$; $m=1,2,\hdots,\frac{N-1}{2}$~\cite{sirker2014boundary}.
\begin{figure}[h!]
\includegraphics[scale=1.15]{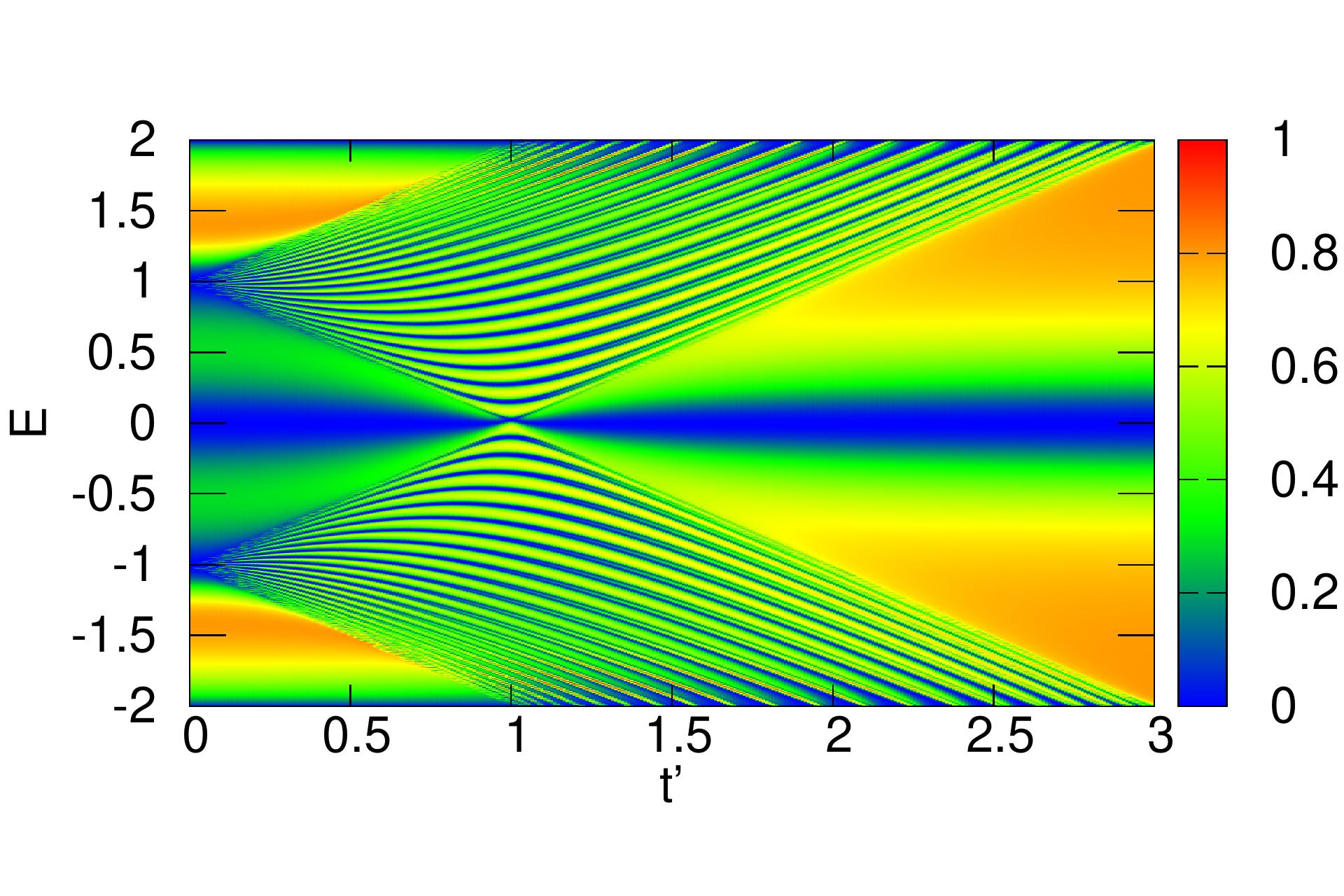}
\caption{Transmission as a function of the incoming wave energy $E$
  for the couplings between the main chain and the open SSH side chain
  with $N=99$ for different inter-cell hopping $t'$. The system
  parameters are set to $v_1=1$, $v_2=1$, $v=1$, $t=1$.
\label{Td_oo}}
\end{figure}
The transmission coefficient as a function of the incoming wave
energies and the inter-cell coupling $t'$ is shown in
\fref{Td_oo}. Full reflection is observed at $E=0$ in both the
regimes, $t'<t$ and $t'>t$. In both the
cases $t'>t$ as well as $t'<t$, the zero energy wave propagates to the
scattering region of the main chain and reflects back (verified from a study of the full system eigenstates shown in
 \fref{wfxn_tbo_2} of the  \aref{wavefn}). This finally results in minimum
transmission through the system. Thus, the narrow line observed
  for $v_2=0$ with $t'<t$ in  \fref{Ts_oo} is modified into a wider
  blue region when $v_2\ne 0$ as shown in  \fref{Td_oo}.

\begin{figure}[h!]
\centering
\includegraphics[scale=0.65]{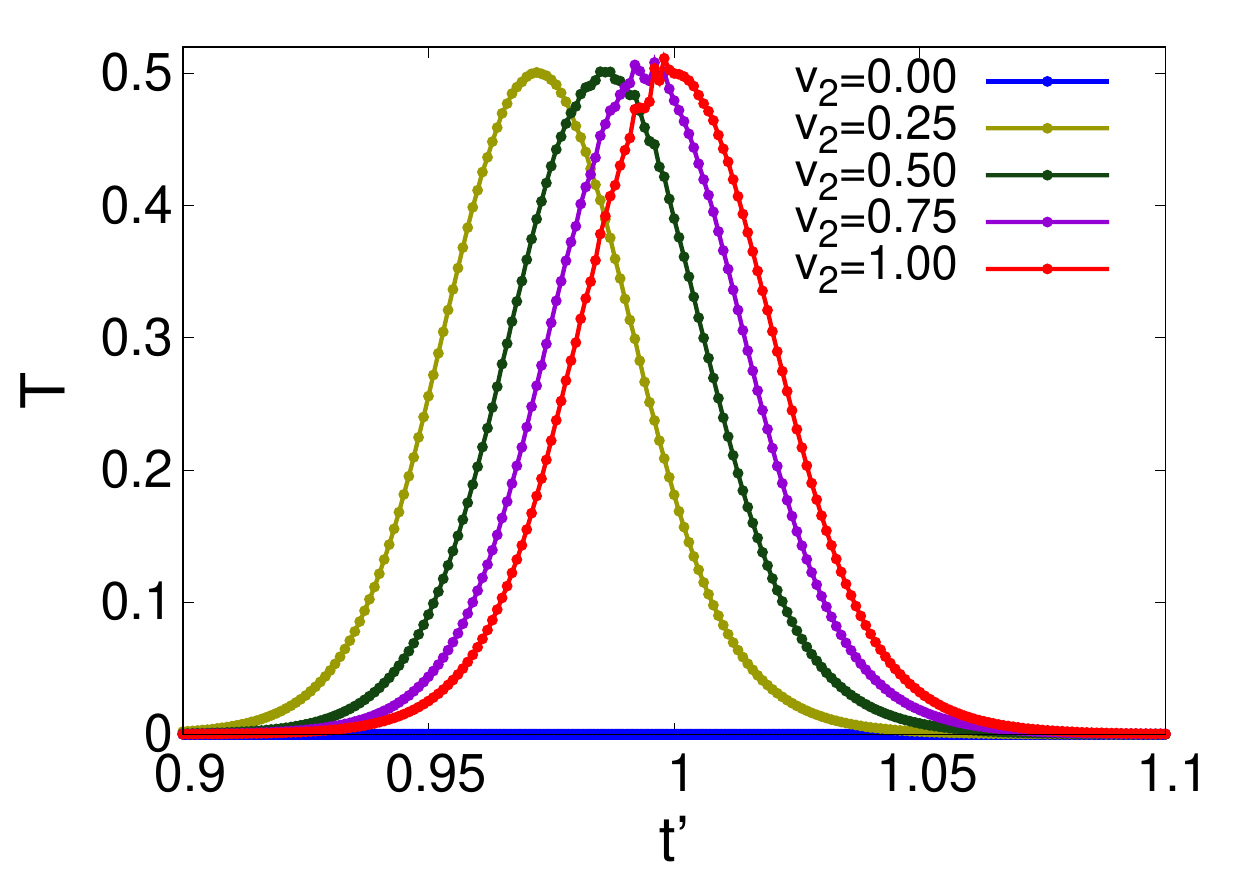}
\caption{The transmission coefficient as a function of different
  inter-cell hopping $t'$ of the open SSH chain for different second
  connection strengths $v_2$. The other system parameters are kept
  fixed: $N=99$, $t=v=v_1=1$, $E\sim0$.}
\label{Td_oo_tdash}
\end{figure}

We have seen that a connection $v_2$ to the side chain significantly
modifies the transmission in the system. This motivates us to study
how tuning the second coupling strength $v_2$ affects transmission for
various inter-leg hopping strengths $t'$ as shown in
 \fref{Td_oo_tdash}. We see that when $v_2=0$, for energy close to
zero, the transmission is zero for any value of $t'$.  However, for
non-vanishing $v_2$ the system shows transmission only in the limit
$t'\rightarrow t$. As $t'$ approaches $t$, the SSH chain starts to
behave as a simple nearest-neighbor tight binding chain and for an odd
system size it should show perfect reflection for energies
$E\sim0$. But due to the presence of the secondary connection the
energy corresponding to perfect reflection (shown in
 \fref{Fano_tbo_1}) is given by \eref{eq22}. Therefore the
system shows some transmission for energies close to zero, as shown in
 \fref{Td_oo_tdash}. In  \aref{lin_chain}, we discuss the
role of the second connection, when the side unit is a simple tight
binding linear chain.


\subsection{Case 2: Open chain with even N}
Next we explore the SSH open chain with an even number of sites as the
side chain. In this case, the isolated SSH model possesses either no
edge state (zero energy state) for the case $t'<t$ or two edge states
located at the two ends for the case $t'>t$ as shown in
\fref{en_ssh}(b).
\begin{figure}[h!]
\includegraphics[scale=1.15]{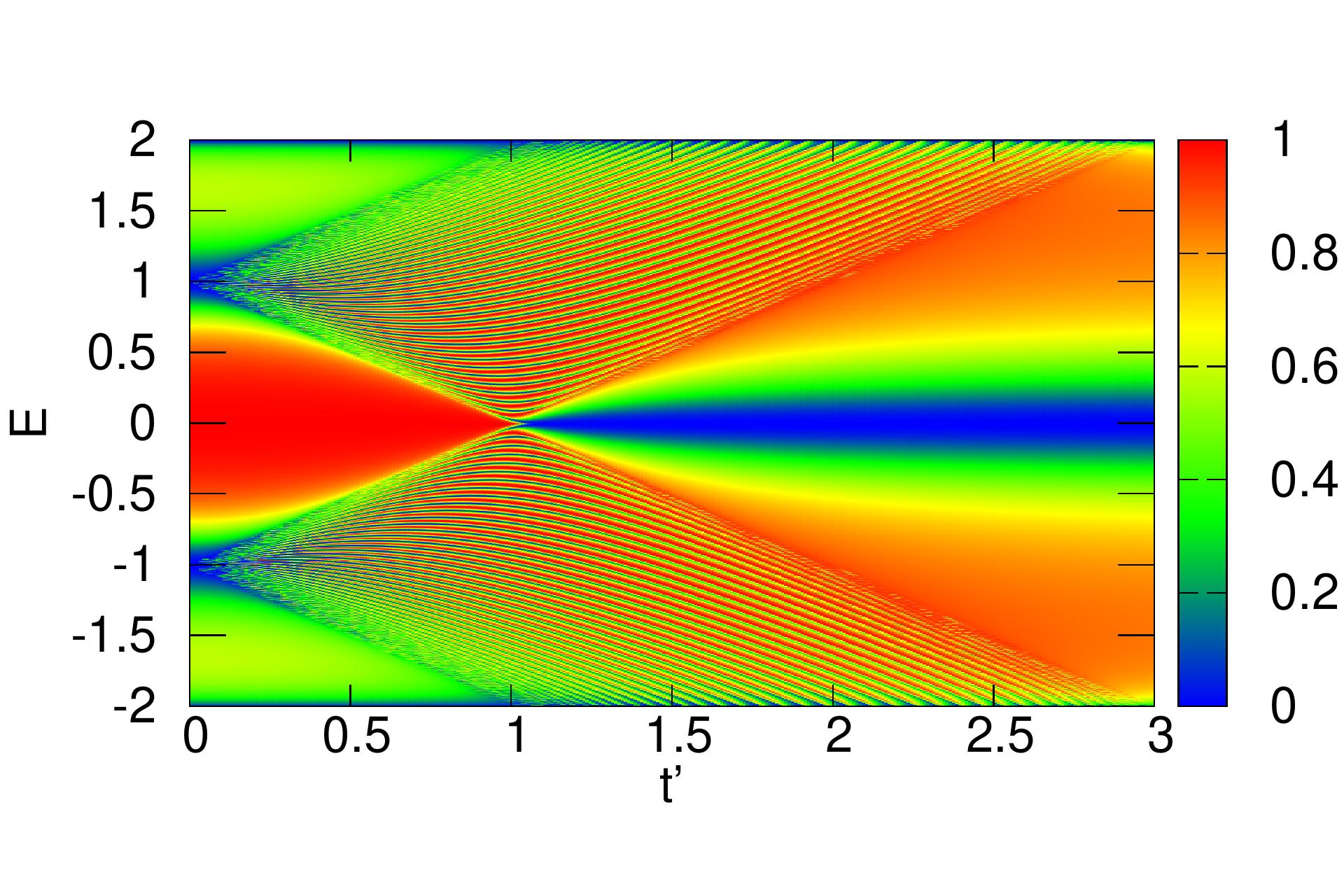}
\caption{The transmission coefficient (T) for a single coupling between the main chain and an open SSH side chain
  with $N=100$ as a function of incoming wave energy
  $E$ and inter-cell hoppings $t'$ with other parameters set to $v=1$, $t=1$, $v_1=1$, $v_2=0$.}
\label{Ts_eo}
\end{figure}
In order to understand the transmission in this system, the required
coefficients of the inverse matrix $\mathcal{H}_N$ are calculated as:
$\alpha_{11}=\alpha_{nn}=\frac{\Gamma_{N-1}^{\{1\}}}{\Gamma_{N}}$,
$\alpha_{1n}=\alpha_{n1}=\frac{(tt')^{\frac{N}{2}}}{t'\Gamma_{N}}$.
In the single coupling setup to the scatting center ($v_1\neq
0;\;v_2=0$), the condition for perfect reflection ($T=0$) is again
given by $\Gamma_N=0$ and the condition for perfect transmission is
given by $\Gamma_{N-1}^{\{1\}}=0$.  The transmission spectrum as
  a function of incoming wave energy $E$ and inter-cell hopping $t'$
  is shown in \fref{Ts_eo}. At $E\sim 0$, the transmission coefficient
  undergoes a transition from $1$ (red region) to $0$ (blue region) on
  switching from the trivial (at $t'<t$) to the topological (at
  $t'>t$) phase of the SSH chain as shown in \fref{Ts_eo}.  
  The propagating wave reflects back from the scattering region in
the main chain as depicted by the
eigenstate of the full Hamiltonian in
\fref{wfxn_tbeo_1} (\aref{wavefn}).

The two links to the side chain modify the condition for
perfect reflection, which now happens for energies satisfying the equation
\begin{equation}
vt'\Gamma_N+v_1v_2(tt')^{\frac{N}{2}}=0,
\label{eq23}
\end{equation}
where
$\Gamma_N=\displaystyle\prod_{\theta_m}(E\pm\sqrt{t^2+{t'}^2+2tt'\cos\theta_m})$. The
variable $\theta_m$ is calculated numerically using the relation
$t\;\Lambda(\theta_m,\frac{N}{2}) +
t'\;\Lambda(\theta_m,\frac{N}{2}-1)=0$, where
$\Lambda(\theta_m,l)=\frac{\sin[(l+1)\theta_m]}{\sin
  \theta_m}$~\cite{sirker2014boundary}.  In the case $t'<t$, the wave
propagates without disturbance in the main chain which results in
perfect transmission in the system.
\begin{figure}[h!]
\includegraphics[scale=1.15]{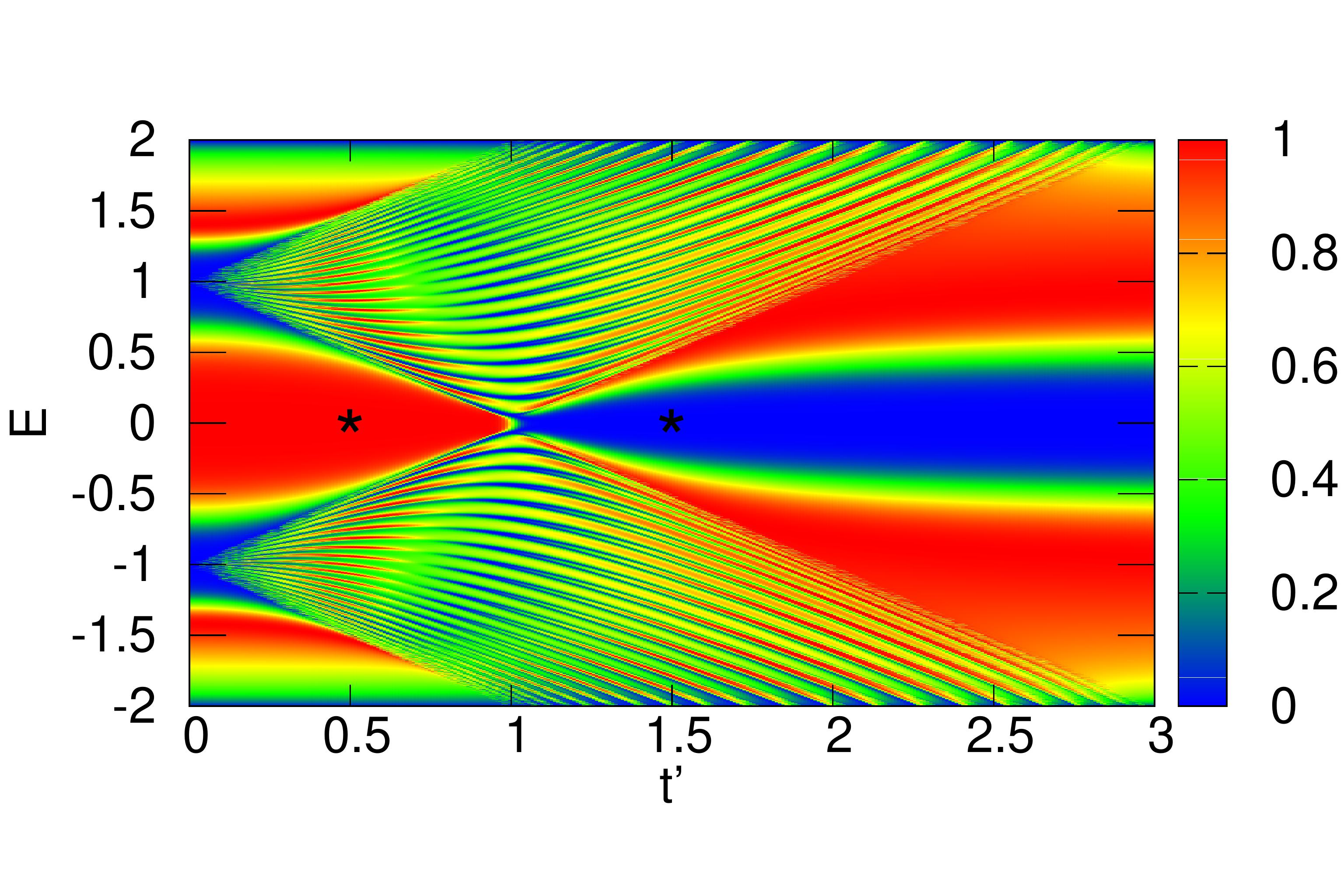}
\caption{The transmission coefficient $T$ for the open SSH side chain
  with length $N=100$ for two couplings ($v_1,\;v_2$) between the main
  chain and the side unit. We show $T$ as a function of inter-cell
  hopping $t'$ and incoming wave energy $E$. The other parameters are
  kept constant: $v=1$, $t=1$, $v_1=1$, $v_2=1$. The wave
    amplitudes corresponding to the two ($*$) symbols are plotted in
    \fref{time_Td_eo}.}
\label{Td_eo}
\end{figure}
However, in the other limit $t'> t$, the main chain is connected to
the two edge states of the SSH chain, and therefore, as shown in
\fref{Td_eo}, minimum transmission is obtained due to the reflection
of the incoming wave by these edge states (\fref{wfxn_tbeo_2}).  Thus,
the transmission coefficient $T$ again features a transition from $1$
to $0$ for zero energy when the side chain undergoes a transition from
the trivial to the topological phase but the width of the perfect
reflection area is increased in the presence of the secondary coupling
$v_2$.  At $E=0$, the transmission coefficient $T$ is unaffected by
the secondary connection to the side chain as shown in \fref{Td_eo}.
This should be seen as a consequence of the robustness of the
topologically protected edge states of the SSH chain.

We verify the results by studying the time dependent wave equation
\begin{equation}
i\frac{d}{d\tau}\psi=H\psi
\label{tdsc}
\end{equation}
where $H$ is the full Hamiltonian of the system, and $\psi$ is the
wavefunction representing wave propagation in the system. The incoming
wave is represented by an initial Gaussian wavepacket (time
$\tau=0$) localized at a site $l_0$ (with $l_0\ll l=1$) which is far
left of the scattering region. It can be written as
\begin{equation}
\psi(j)=e^{-\iota k j}e^{-\frac{1}{2}\big(\frac{j - l_0}{\sigma}\big)^2},
\label{eq:19}
\end{equation}
where $\sigma$ is the width and $k$ is the momentum. The  wavepacket at a later time $\tau_0$ can be obtained from \eref{tdsc} as
\begin{equation}
\psi(j,\tau_0)=e^{-\iota H \tau_0}\psi(j,0).
\end{equation}
 The initial wavepacket evolves in time and moves towards the
 scattering region. The post-scattering features of the wavepacket are
 governed by the Hamiltonian of the side unit.  We focus on the wave
 amplitude in the system after the scattering when the energy of the
 incoming wavepacket is $E=0$ at two points indicated by star ($*$)
 symbols in  \fref{Td_eo}.
\begin{figure}[h!]
\includegraphics[scale=0.561]{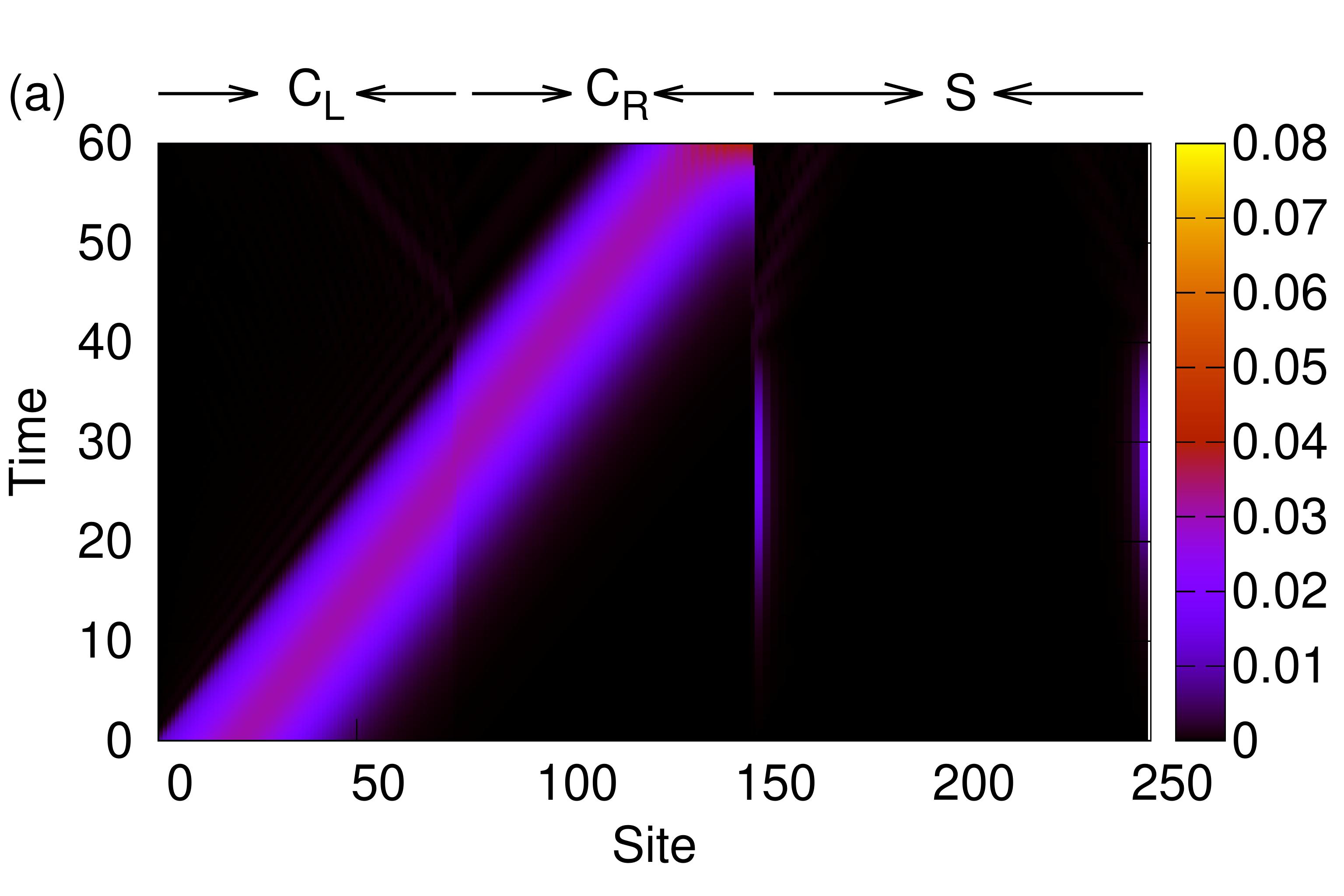}
\includegraphics[scale=0.561]{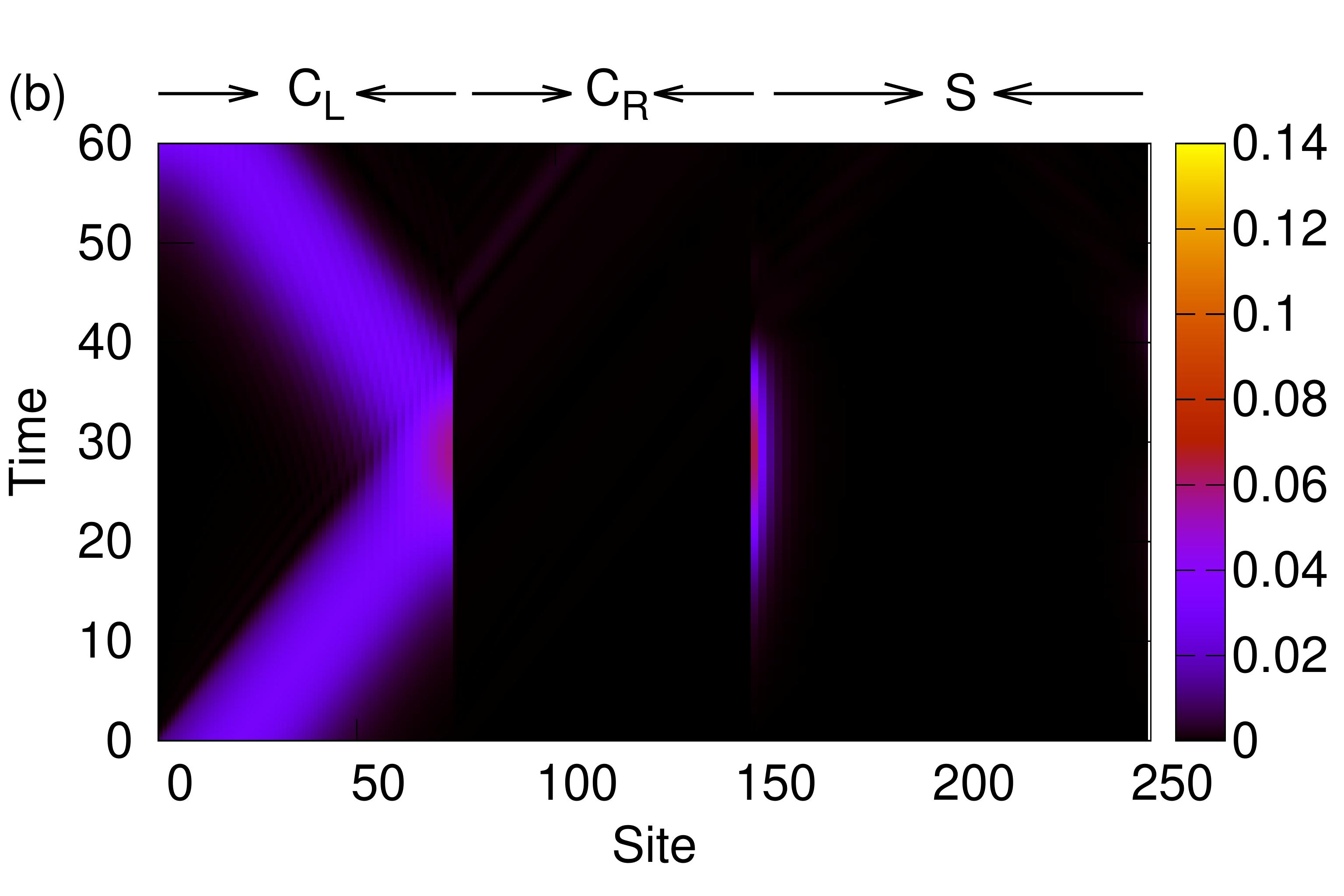}
\caption{The evolution of the square of the wave amplitude for an
  incoming wave-packet with energy $E=0$ in the system. The main chain
  ($site=1,2,\hdots,N_c$) with hopping $v=1$ and length $N_c=150$ is
  connected to an open SSH side chain ($site=N_c+1,\hdots,N_c+N$) of
  length $N=100$ with couplings $v_1=1$, $v_2=1$. We mark the left and
  right parts of the main chain as $C_L$ and $C_R$ and the SSH
  side-unit as $S$ for clarity. Here $t=1$. (a) For $t'=0.5$ we find
  perfect transmission and for (b) $t'=1.5$ perfect reflection. The
  wave amplitudes correspond to the two ($*$) symbols shown in
  \fref{Td_eo}.}
\label{time_Td_eo}
\end{figure} 
 In  \fref{time_Td_eo}(a) and (b) we show the square of the wave
 amplitude for $t'=0.5$ and $t'=1.5$, respectively. In the trivial
 phase ($t'<t$), the entire wave-packet is transmitted beyond the
 scattering region indicating complete transmission. In the
 topological phase $t'>t$ on the other hand, the wave packet is
 completely reflected from the scattering region. Thus, the dynamics
 reaffirms the results obtained from TMM, despite the finite energy width of the wavepacket \eqref{eq:19}. This is enabled by the broad spectral width of the total reflection and total transmission features in  \fref{Td_eo} for $t'<t$ and $t'>t$ respectively.

\begin{figure}[h!]
\centering
\includegraphics[scale=0.6]{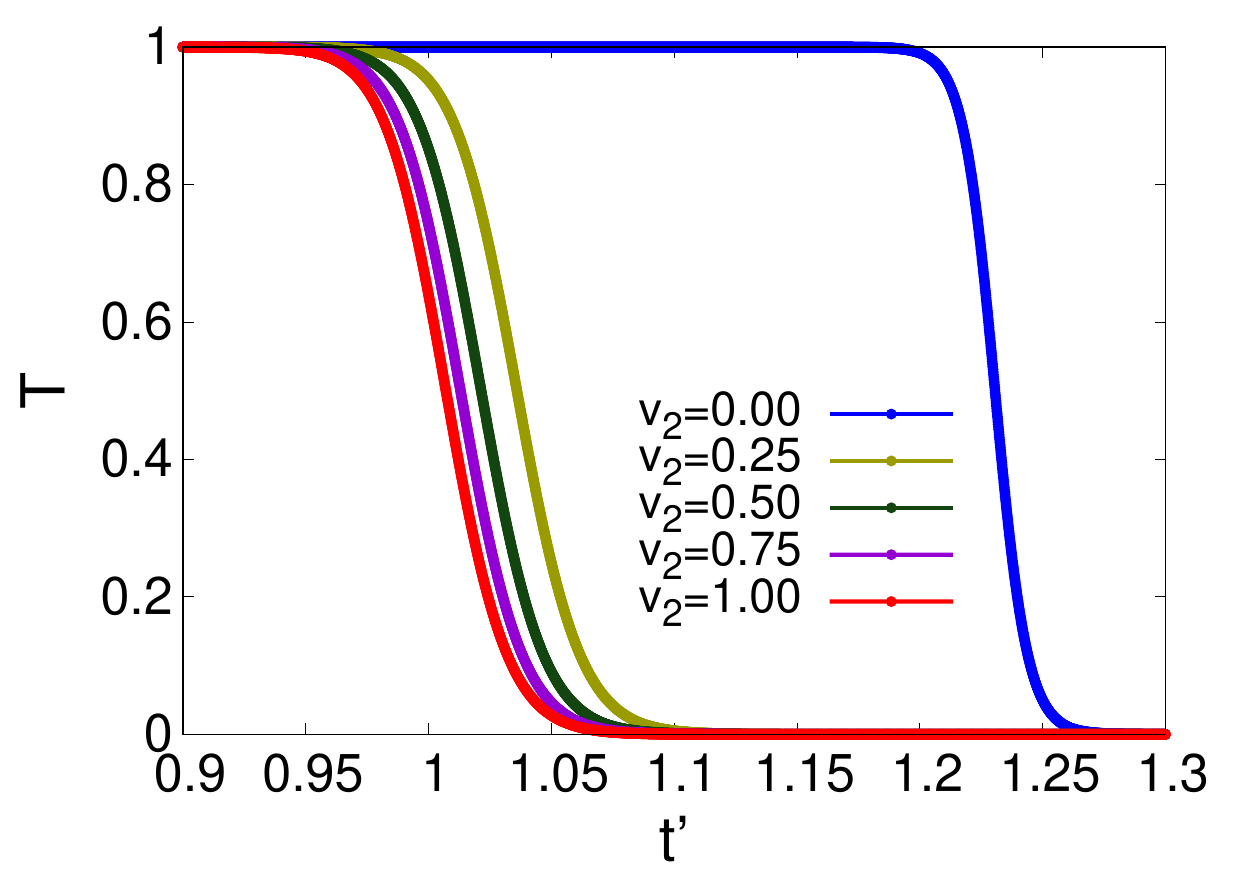}
\caption{The transmission coefficient as a function of different
  inter-cell hopping $t'$ of the open SSH side chain for different
  second connection strengths $v_2$ with other system parameters fixed
  at $N=100$, $t=v=v_1=1$, $E\sim 0$.}
\label{Td_eo_tdash}
\end{figure}
Finally, we show the transmission coefficient as a function of $t'$
for different values of $v_2$ in \fref{Td_eo_tdash}.  The transmission
coefficient shows a step-like drop from $1$ to $0$ as the system
switches from the trivial to the topological phase at $t'=t$. Usually,
this phase transition happens exactly at $t'=t$ only for large system
sizes. However, \fref{Td_eo_tdash} shows that the secondary connection
facilitates this phase transition as the jump in $T$ at $t'=t$ becomes
sharp even for small system sizes. The combination of an energetically
broad region of total reflection (for $t'<t$) and total transmission
(for $t'>t$) is evident in \fref{Td_eo}, with a fairly abrupt
transmission between the two in terms of $t'$ as seen in
\fref{Td_eo_tdash}, might have useful switching applications for
devices.


\subsection{Case 3: Closed chain with even N}
In this subsection we look at the last case, namely the closed SSH
chain with an even number of sites. In this scenario, the ends of the
side chain are connected to the main chain as well as to each other as
shown in \fref{FA_ssh}. The closed isolated SSH chain becomes gapless
at $t'=t$ with the two energy bands touching at zero energy as
depicted in \fref{en_ssh}(c).  This results in non-zero energy states
in both regions $t'<t$ and $t'>t$.
\begin{figure}[h!]
\includegraphics[scale=1.15]{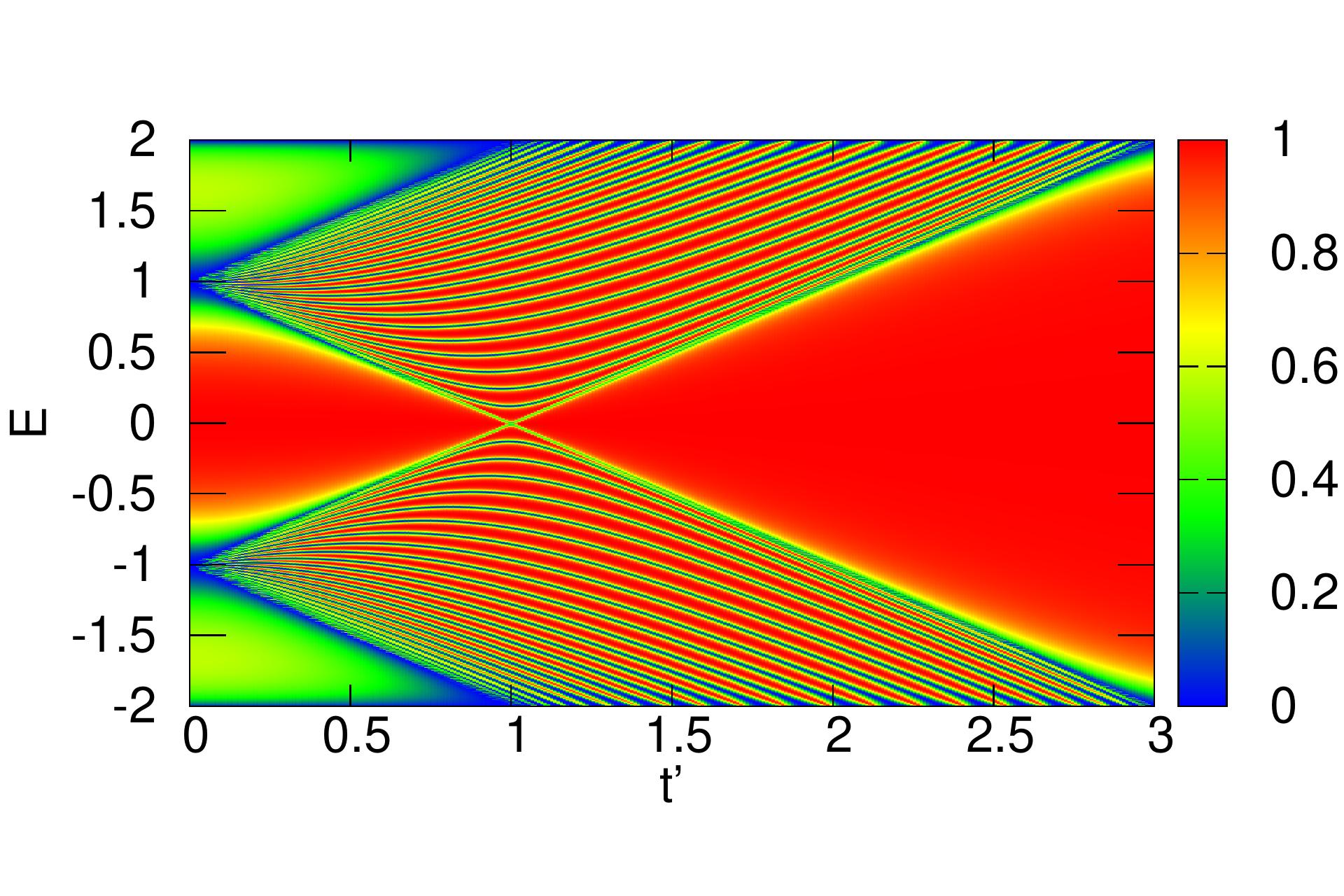}
\caption{The transmission (T) in the system for various incoming wave
  energies $E$ and inter-cell couplings $t'$ when a single connection
  exists between the main unit and a closed SSH side unit with length
  $N=100$. The other parameters are kept constant: $v=1$, $t=1$,
  $v_1=1$, $v_2=0$.}
\label{Ts_ep}
\end{figure}
Here, the coefficients of the inverse of the matrix $\mathcal{H}_N$
are given by
$\alpha_{11}=\alpha_{nn}=\frac{\Gamma_{N-1}^{\{1\}}}{\Gamma_{N}}$,
$\alpha_{1n}=\alpha_{n1}=\frac{(tt')^{\frac{N}{2}}+{t'}^2\Gamma_{N-2}^{\{1,N\}}}{t'{\Gamma_{N}}}$.
The relation
$\Gamma_{N}=\displaystyle\prod_{\theta_m}(E\pm\sqrt{t^2+{t'}^2+2tt'\cos\theta_m})$
with $\theta_m=\frac{4m\pi}{N}$; $m=1,2,\hdots,N/2$ yields
$\Gamma_{N}$~\cite{sirker2014boundary}.  Figure ~\ref{Ts_ep} shows the
behavior of the transmission coefficient $T$ of the system in the
presence of a single connection to the side chain in the two phases
($t'>t$ and $t'<t$) of the system.  Again, perfect transmission is
attained for the incoming energies satisfying $\Gamma_{N-1}^{\{1\}}=0$
whereas perfect reflection is attained for energies consistent with
$\Gamma_{N}=0$ as shown in \fref{Ts_ep}. The system shows full
transmission for energies close to zero in both the regions ($t'>t$
and $t'<t$) of the SSH chain due to the opening of the energy bands.

\begin{figure}[h!]
\includegraphics[scale=1.15]{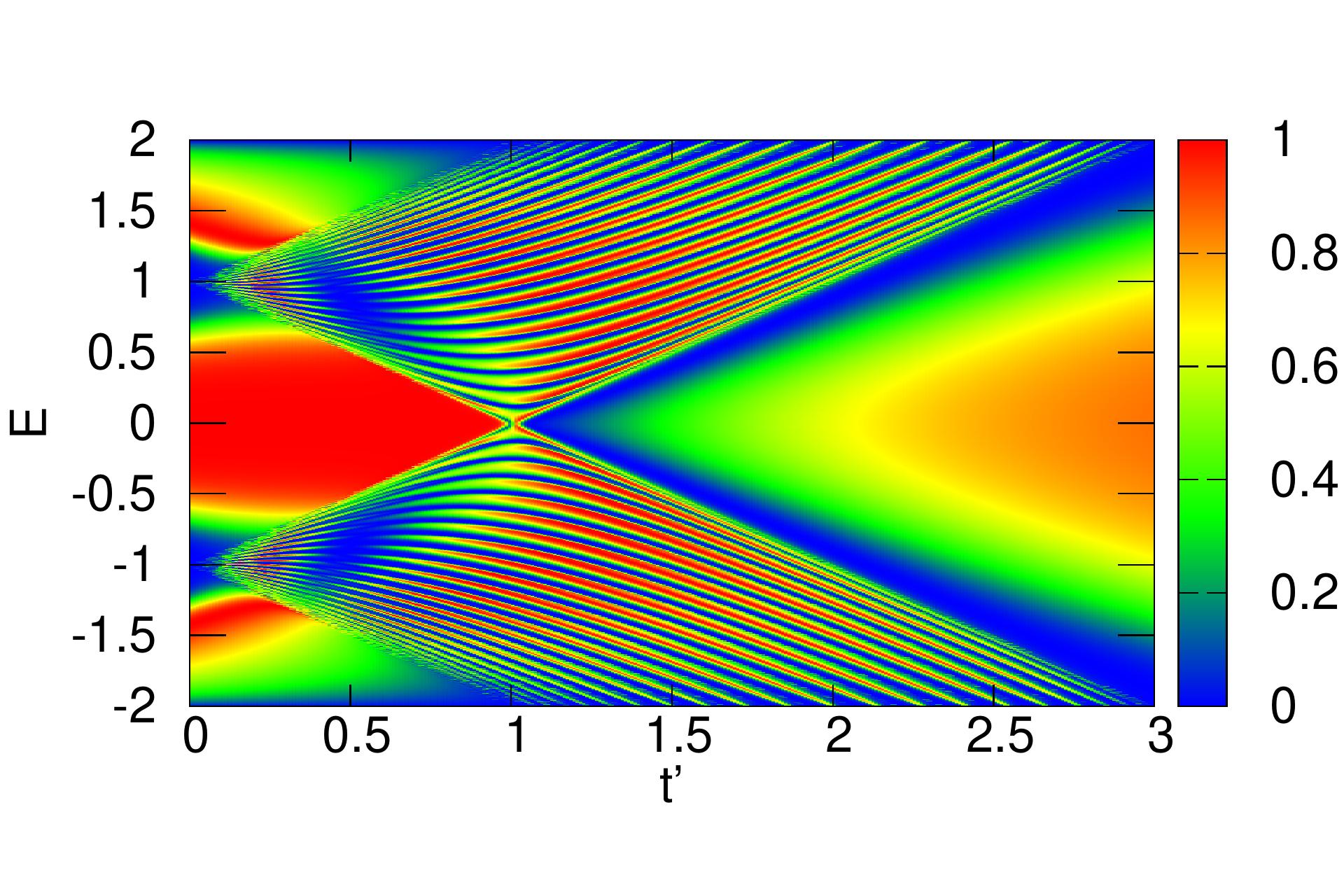}
\caption{The transmission coefficient $T$ for the closed SSH side
  chain with $N=100$ for double coupling between the main chain and
  the side unit. We show $T$ as a function of inter-cell hopping $t'$
  and incoming wave energy $E$ with other fixed system parameters as
  $v=1$, $t=1$, $v_1=1$, $v_2=1$.}
\label{Td_ep}
\end{figure}

The addition of one more connection ($v_2$) further modifies the condition
for perfect reflection which now happens for energies consistent with:
\begin{equation}
v\Gamma_{N}+v_1v_2t'\Gamma_{N-2}^{\{1,N\}}+v_1v_2(t)^{\frac{N}{2}}(t')^{\frac{N}{2}-1}=0.
\label{eq25}
\end{equation}
The behavior of the transmission coefficient in the presence of the
two adjacent connections to the side chain is shown in \fref{Td_ep}.
The transmission profile exhibits a change from perfect transmission
  to full reflection in the vicinity of the trivial-to-topological
  point as featured in \fref{Td_ep}. In the extreme case
  $t'\ll t$, the system shows full transmission whereas in the other
  regime when $t'>t$, a suppression of transmission with
  increasing second coupling strength ($v_2$) is seen due to modified
  reflection conditions given by \eref{eq25}.

\begin{figure}[h!]
\centering
\includegraphics[scale=0.7]{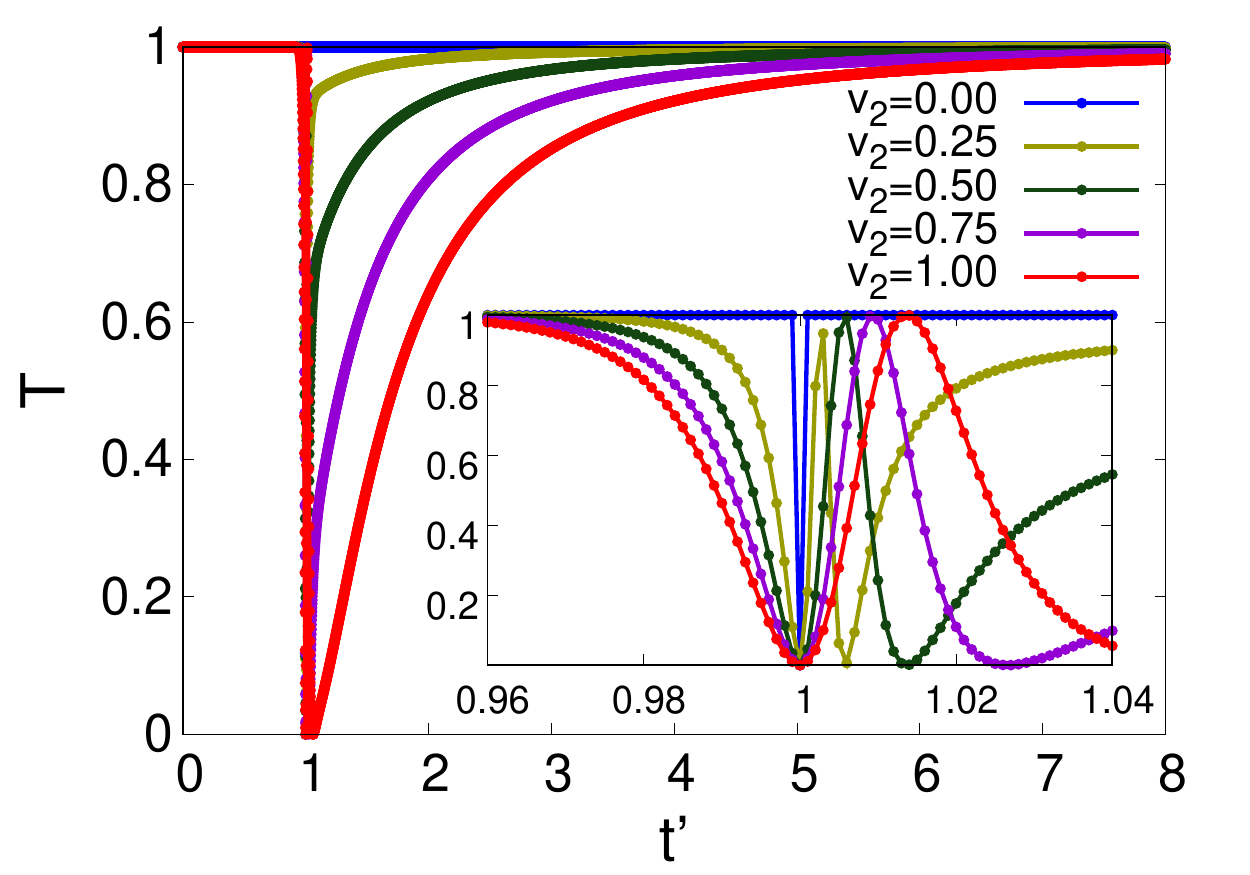}
\caption{The transmission coefficient as a function of different
  inter-cell hopping $t'$ of the closed SSH side chain for different
  second connection strengths $v_2$. The other system parameters are
  kept fixed at: $N=100$, $t=v=v_1=1$, $E\sim0$. The inset is a
  zoomed version that highlights the intricate structure in the
  vicinity of the phase transition point.}
\label{Td_ep_tdash}
\end{figure}
The behavior of the transmission for a range of secondary connection
strengths for different inter-leg coupling $t'$ is depicted in
\fref{Td_ep_tdash}. In the absence of the second connection $v_2$ the
transmission coefficient shows a sharp dip at the transition point
$t'=t$. However, in the presence of the second connection $v_2$ the
width of the dip at $t'=t$ is split into a dip and a peak in the
vicinity of the transition point $t'=t$. This is due to the extra
connection to the main chain which modifies the transmission
conditions in the system.


\section{\label{exp}Experimental Realization} 
  In this section, we briefly discuss the possible experimental
  realization of our setup. One way to generate the Fano-Anderson
  chain structures discussed here is using optical lattices filled
  with ultra cold atoms like Rubidium
  ($^{87}Rb$)~\cite{theis2005optical,krinner2017two,gross2017quantum}.
  For strong lattices, these realize a tight binding chain for the
  atoms, while when replacing direct tunneling with laser assisted
  tunneling between some sites, alternating tunneling couplings as
  required here can be
  realized~\cite{PhysRevLett.111.185302,PhysRevLett.111.185301}. This
  technique can additionally be combined with employing
  superlattices~\cite{Gerbier_2010}.
  
  A second platform where our results are directly applicable is
  photonic crystals~\cite{inoue2004photonic,Yu12743}. Photonic crystal
  technology is based on the creation of nanocavities with a photonic
  band gap~\cite{kuramochi2016manipulating,busch2003wannier} which are
  introduced into a photonic crystal waveguide. These exclusion zones
  can then practically mimic a tight binding chain for photons, where
  control over the coupling strengths can be exerted by adjusting the
  spacing between the sites. Recently, an SSH model with a combination
  of rigid and elastic materials, which could continuously be tuned
  across its topological phase transition by stretching
  ~\cite{naz2018topological} was proposed. By immersing the photonic
  crystal into a nematic liquid crystal background, dynamical
  tunability with the help of an external electric field could also be
  achieved ~\cite{shalaev2018reconfigurable}.


\section{\label{sec5}Conclusion}
In this work, we analyse the interplay of Fano resonances, topology and
edge states in a Fano-Anderson chain possessing a topological side
unit. We take the SSH chain which possesses topological
characteristics as a prototype side unit, and show that the topology of the
side unit modifies the Fano resonance profile and thus the transport
probability past the topological unit.  We observe that the
transmission profile is modified when the system is tuned from the
trivial to the topological phase. With a single connection between the
main chain and the side unit, the topological to trivial transition
cannot be efficiently detected. However, if two connections are present, a
clear signal of the transition is obtained directly from the
transmission profile. We provide a detailed study of how the
transmission profile depends on the boundary conditions and the
topological properties of the SSH chain. The detection of topological
characteristics from Fano resonances in quantum transport could allow
a new experimental handle on topological states.

We explicitly derive an exact expression for the transmission
coefficient using the transfer matrix method and obtain conditions for
Fano resonance as well as perfect reflection. The expression obtained
can be generalized to an arbitrary side unit and hence may find
application in other studies of Fano resonance assisted
transport. This enables us to show that an open topological chain with
an even number of sites exhibits a sharp transition between complete
reflection or complete transmission of all waves with energy near
zero, depending on the parameters of the topological scatterer. The
sharpness of this transition paired with its wide bandwidth should
make this a useful feature for switching in device applications
involving photonic crystals or nano-materials with topological
elements.

\section*{Acknowledgments}
A.S acknowledges financial support from SERB via the grant (File
Number: CRG/2019/003447), and from DST via the DST-INSPIRE Faculty
Award [DST/INSPIRE/04/2014/002461].


\bibliographystyle{unsrt} \bibliography{ref}


\appendix

\section{Transmission coefficient}
\label{TMM}
We assume that an incoming wave from the far left approaches the
scattering region in the main chain and that after the scattering
time period, the wave has been split in two parts in the main chain; a
reflected part moving towards the left and a transmitted part moving
towards the right as shown in  \fref{structure}(b). The boundary
conditions for the stationary scattering states in the time independent
scattering problem can be written as:
\begin{equation}
A_\ell = I_0 e^{\iota k\ell} + r e^{-\iota k\ell}
\label{bc_l}
\end{equation}
for $\ell<1$ and
\begin{equation}
A_\ell = t_{out} e^{\iota k\ell}
\label{bc_r}
\end{equation}
for $\ell>2$. The coefficients corresponding to the incoming, reflected
and transmitted waves are denoted respectively as $I_{0}$, $r$, and
$t_{out}$. We can now formulate a transfer matrix that connects the wavefunction on nearby sites of the main chain represented by the lattice equation
 \eref{latt_eqnfinal} as:
\begin{equation}
\begin{bmatrix}
&\hspace{-7pt}A_{\ell+1}\\
&\hspace{-7pt}A_{\ell}
\end{bmatrix}=
M_T
\begin{bmatrix}
&\hspace{-7pt}A_{\ell}\\
&\hspace{-7pt}A_{\ell-1}
\end{bmatrix} ,
\label{tmm}
\end{equation}
where $M_T$ is the transfer matrix given by
\begin{equation}
M_T=\begin{bmatrix}
&\frac{E-\alpha_{pp}v^{2}_{1}\delta_{\ell,1}-\alpha_{qq}v^{2}_{2}\delta_{\ell,2}}{v+v_1v_2\alpha_{pq}\delta_{\ell,1}} &-\frac{v+v_{2}v_{1}\alpha_{qp}\delta_{\ell,2}}{v+v_1v_2\alpha_{pq}\delta_{\ell,1}}\\
&1 &0
\end{bmatrix}.
\label{trans_matr}
\end{equation}
As depicted in  \fref{structure}, the side unit and main chain are
linked by two connections with different hopping $v_1$ and $v_2$. Hence, we can represent the
final transfer matrix of the system as a product of two transfer
matrices at the two connecting sites.  The first transfer matrix for
the connection at the first site $\ell=1$ is given by $M_1=(M_T)_{\ell=1}$
and the second transfer matrix for the connection at the next site
$\ell=2$ is given by $M_2=(M_T)_{\ell=2}$.  Thus, the final transfer matrix
is given by $M=M_2 M_1$ which connects wavefunctions of the parts that
lie to the left and right of the scattering regions as
\begin{equation}
\begin{bmatrix}
&A_{3}\\
&A_{2}
\end{bmatrix}=
M\begin{bmatrix}
&A_{1}\\
&A_{0}
\end{bmatrix} =
M_2 M_1\begin{bmatrix}
&A_{1}\\
&A_{0}
\end{bmatrix}
\end{equation}
with
\begin{equation}
\label{eq:23}M=\begin{bmatrix}
&M_{11} &M_{12}\\
&M_{21} &M_{22}
\end{bmatrix},
\end{equation}
where $M_{ij}=\frac{M^{'}_{ij}}{v(v+V')}$ with $M^{'}_{11}=[(E-V_{1})(E-V_{2})-|v+V'|^2]$, $M^{'}_{12}=-v(E-V_{2})$, $M^{'}_{21}=v(E-V_{1})$, $M^{'}_{22}=-v^2$.
Thus, the final transfer matrix for the setup is given by $M$.

The transmission coefficient $T = |t^2_{out}/I_0^2|$ is calculated from the transfer matrix using the boundary conditions (\eref{bc_l} and \eref{bc_r}) as:
\begin{equation}
T=\frac{4\sin^2k}{|M_{11}e^{-ik}+M_{12}-M_{21}-M_{22}e^{ik}|^2}.
\label{trans_1}
\end{equation}
Also, the transmission coefficient in terms of the energy of the incoming wave can be written as: 
\begin{equation}
T=\frac{v^2(4v^2-E^2)|v+V'|^2}{[\beta v^2+\gamma v E-E^2 M^{'}_{11} M^{'}_{22}]},
\label{trans_2_1}\end{equation}
where $\beta=(M^{'}_{11}+M^{'}_{22})^2+(M^{'}_{12}-M^{'}_{21})^2$ and $\gamma=(M^{'}_{11}-M^{'}_{22})(M^{'}_{12}-M^{'}_{21})$.

\section{Wavefunctions for the edge states}
\label{wavefn}
Here, we discuss the eigenstates of the complete Hamiltonian which
incorporates the main chain as well as the side SSH chain. The
properties of a wave travelling in the main chain with a particular
energy are intimately connnected to these eigenstates of the complete
Hamiltonian. We focus on the eigenstates corresponding to the energies
at which a Fano resonance dip is observed in the transmission profile.

\begin{figure}[h!]
\includegraphics[scale=0.168,angle=-90]{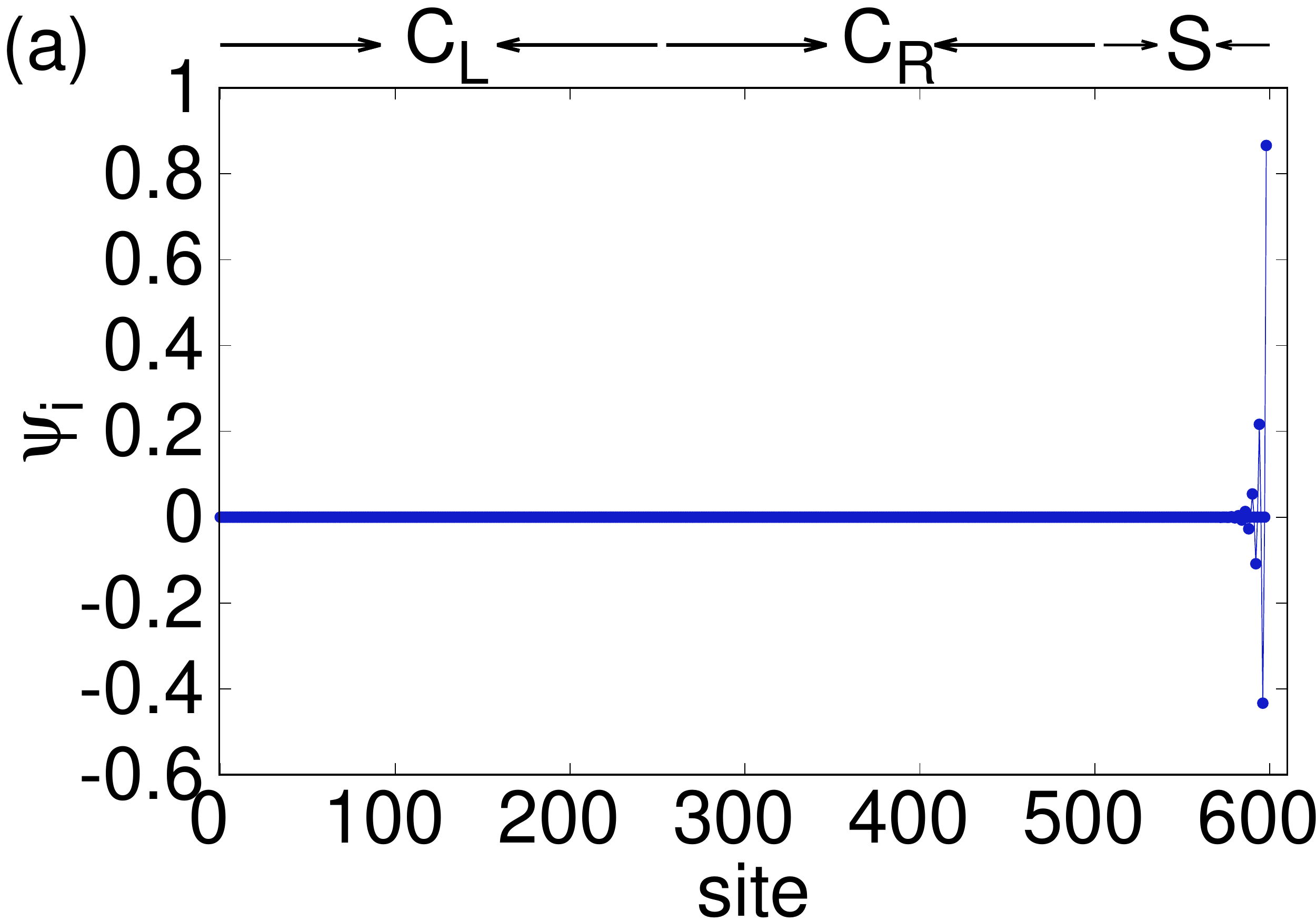}
\includegraphics[scale=0.168,angle=-90]{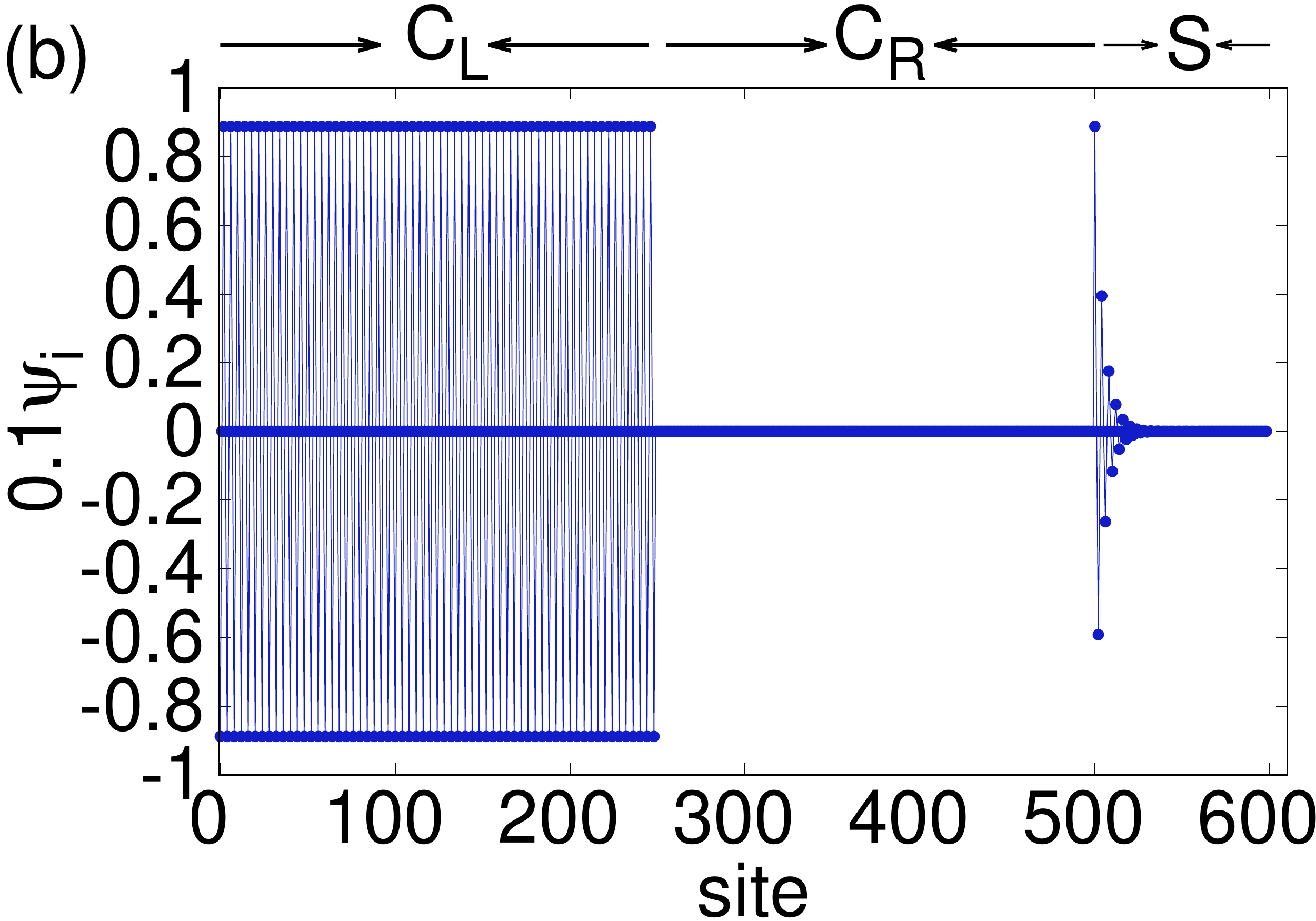}
\caption{The zero energy wavefunction of the system when the main
  chain is the simple tight binding chain with length $N_c=500$ (corresponding to site
  indices running from $1$ to $500$) to which an open SSH side chain
  of length $N=99$ (whose site indices run from $501$ to $599$) is
  connected to one of the two middle sites. We mark the left and
  right parts of the main chain as $C_L$ and $C_R$ and the SSH
  side-unit as $S$ for clarity.  The connecting hopping
  strengths are $v_1=1.0$, $v_2=0$ and the other parameters as
  $v=1.0$, $t=1.0$ with (a) $t'=0.5$ and (b) $t'=1.5$.}
\label{wfxn_tbo_1}
\end{figure}

\fref{wfxn_tbo_1} features the edge states of the system with a single
connection ($v_1\neq0$ and $v_2=0$) between the main chain and the SSH
chain possessing an odd number of sites.
In the region $t'<t$, the edge state of the isolated SSH chain lies
predominantly on the edge site of the chain and the probability
amplitude decays towards the bulk. The state at zero energy of the
full system also shows the same behavior as depicted in
\fref{wfxn_tbo_1}(a). The eigenstate covers only a few sites close to
the free end of the SSH chain and no probability amplitude is seen
in the main chain. We would therefore expect no contribution to
transport from this eigenstate and thus a very sharp dip is observed
in the transmission profile at $E=0$. The eigenstates close to $E=0$
are completely different as they possess probability amplitudes on the
main chain (both towards the left and right of the defect) and allow a
finite amount of transmission (\fref{Ts_oo}).

 In the region $t'>t$, the edge state mainly lies on the connected end
 of the SSH chain and decays in the bulk of the SSH chain. The
 connected end shows edge-state character here, despite its connection
 to the main chain. However, the amplitude at this end of the SSH
 chain is much smaller than in the other case in \fref{wfxn_tbo_1}(a)
 which is similar to the the edge state of the isolated SSH chain.
 Furthermore, as shown in \fref{wfxn_tbo_1}(b), the eigenstate for the
 full system also shows a finite probability amplitude in the main
 chain that is to the left of the defect. We take an even total number
 of sites in the main chain so that the number of sites to
 the left of the defect is different from the number of sites to the
 right of the defect, thus breaking the left-right symmetry.  As the
 eigenstate lies merely on one side of the main chain, zero
 transmission at this energy is observed as depicted in \fref{Ts_oo}.
 
\begin{figure}[h!]
\includegraphics[scale=0.168,angle=-90]{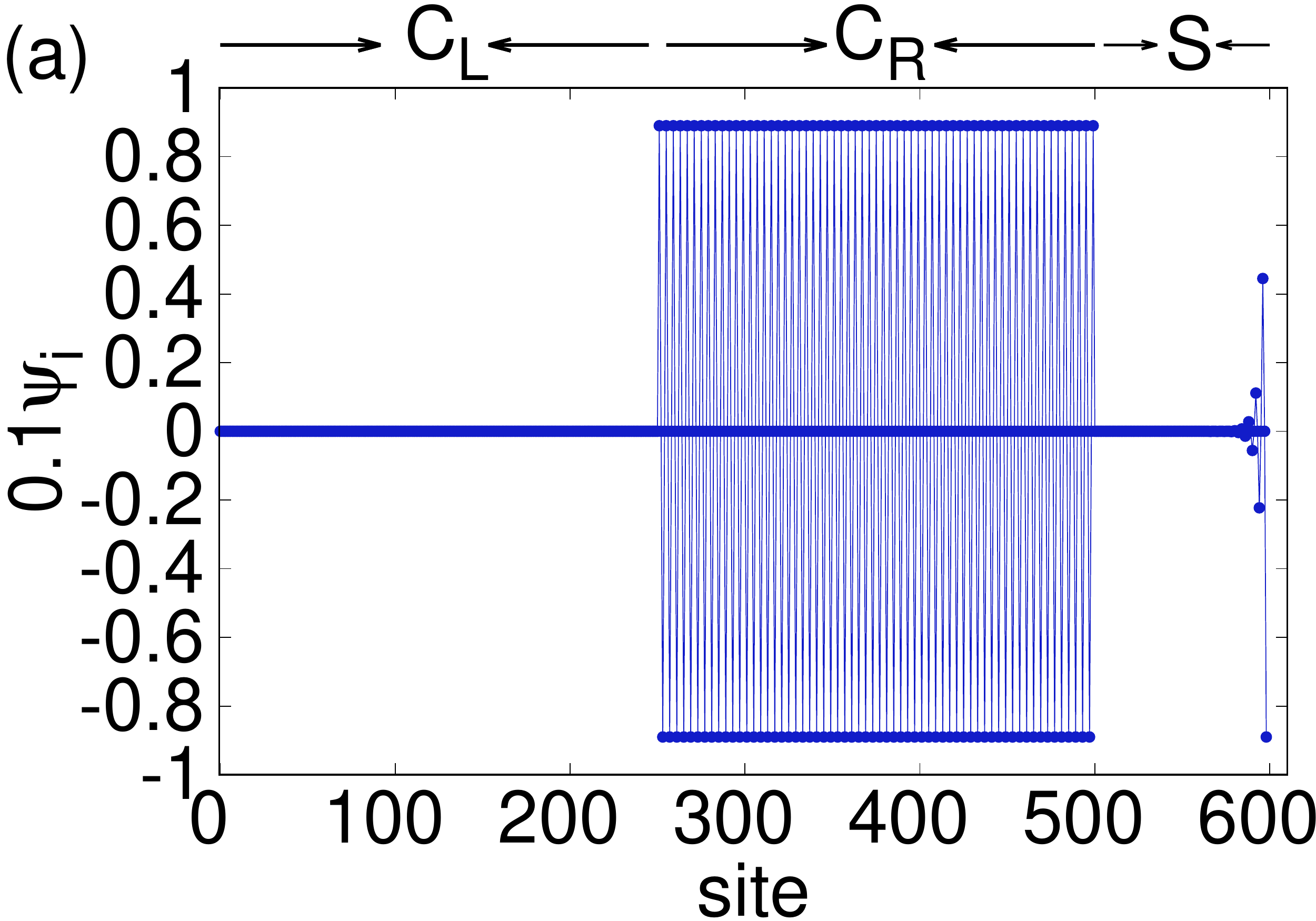}
\includegraphics[scale=0.168,angle=-90]{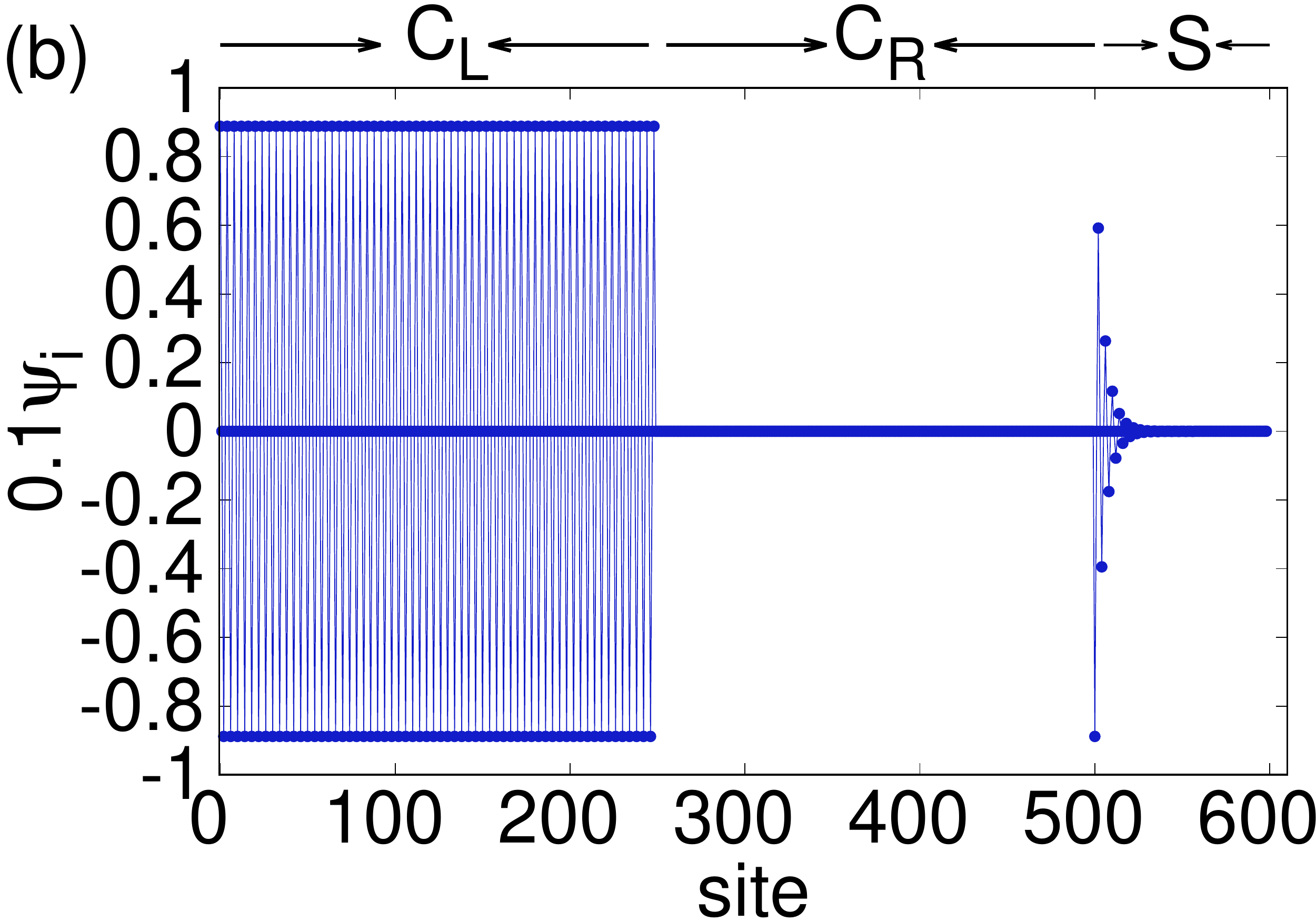}
\caption{The zero energy wavefunction of the system when the main
  chain is the simple tight binding chain with $N_c=500$ (site indices
  running from $1$ to $500$) to which an open SSH chain of length
  $N=99$ (whose site indices run from $501$ to $599$) is connected
  to one of the middle sites and an adjacent site. The connecting
  hopping strengths are $v_1=1.0$, $v_2=1.0$ and the other parameters
  are $v=1.0$, $t=1.0$, and (a) $t'=0.5$ and (b) $t'=1.5$.}
\label{wfxn_tbo_2}
\end{figure}

Next, we explore the eigenstate of the full system with two
connections ($v_1\neq0,\;v_2\neq0$) between the side chain and the
main chain ( \fref{wfxn_tbo_2}).
In \fref{wfxn_tbo_2} we show the eigenstate corresponding to $E=0$ for
both cases $t'>t$ and $t'<t$. In both cases, the eigenstate of the
full system vanishes in one half of the main chain. As a consequence
of this, again the transmission in the system is fully suppressed as
shown in \fref{Td_oo}. With the double coupling, both the ends of the
SSH chain are connected to the main chain. We observe that a finite
probability amplitude is seen at (a different) one of the edges in the
two phases.

\begin{figure}[h!]
\includegraphics[scale=0.168,angle=-90]{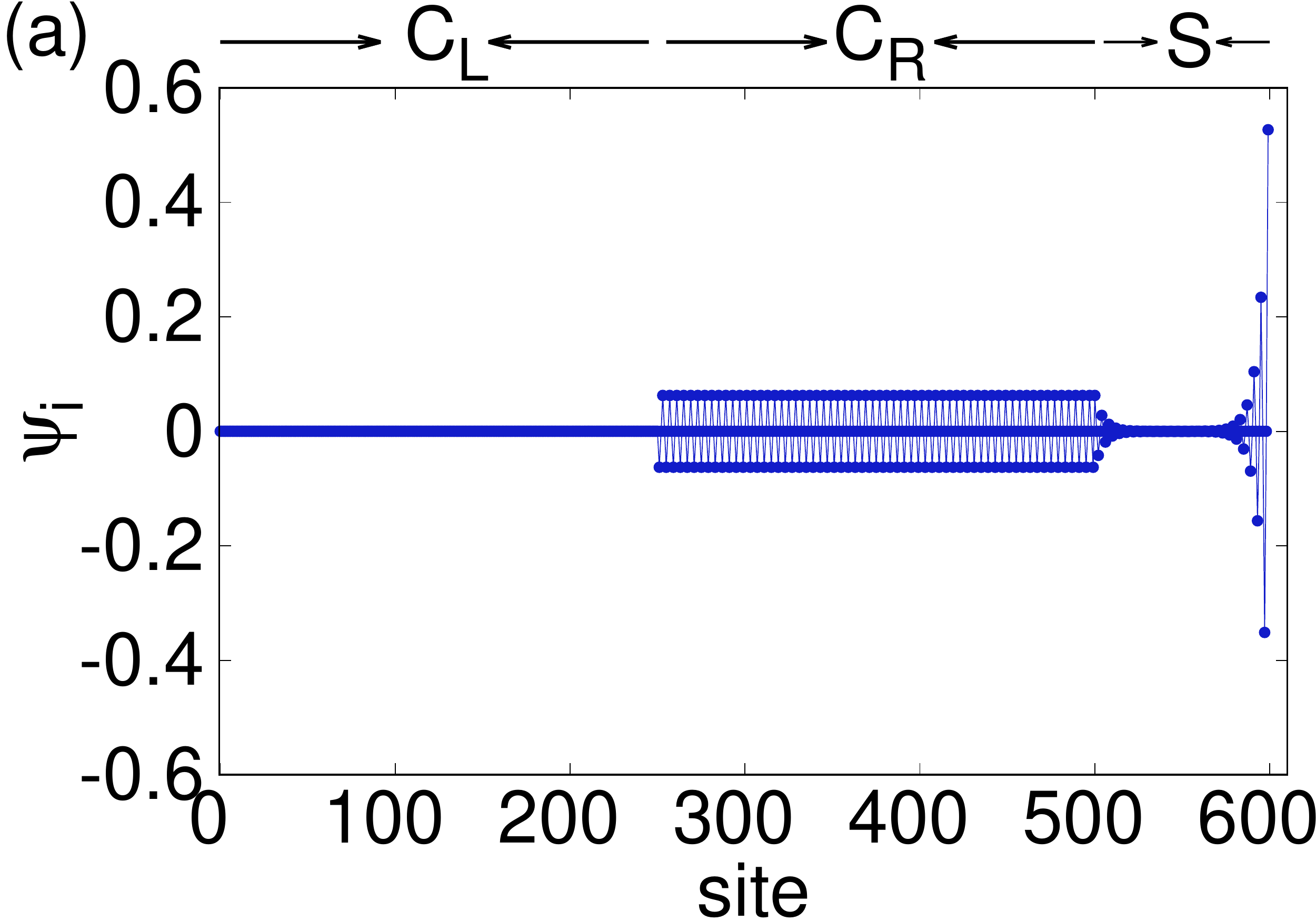}
\includegraphics[scale=0.168,angle=-90]{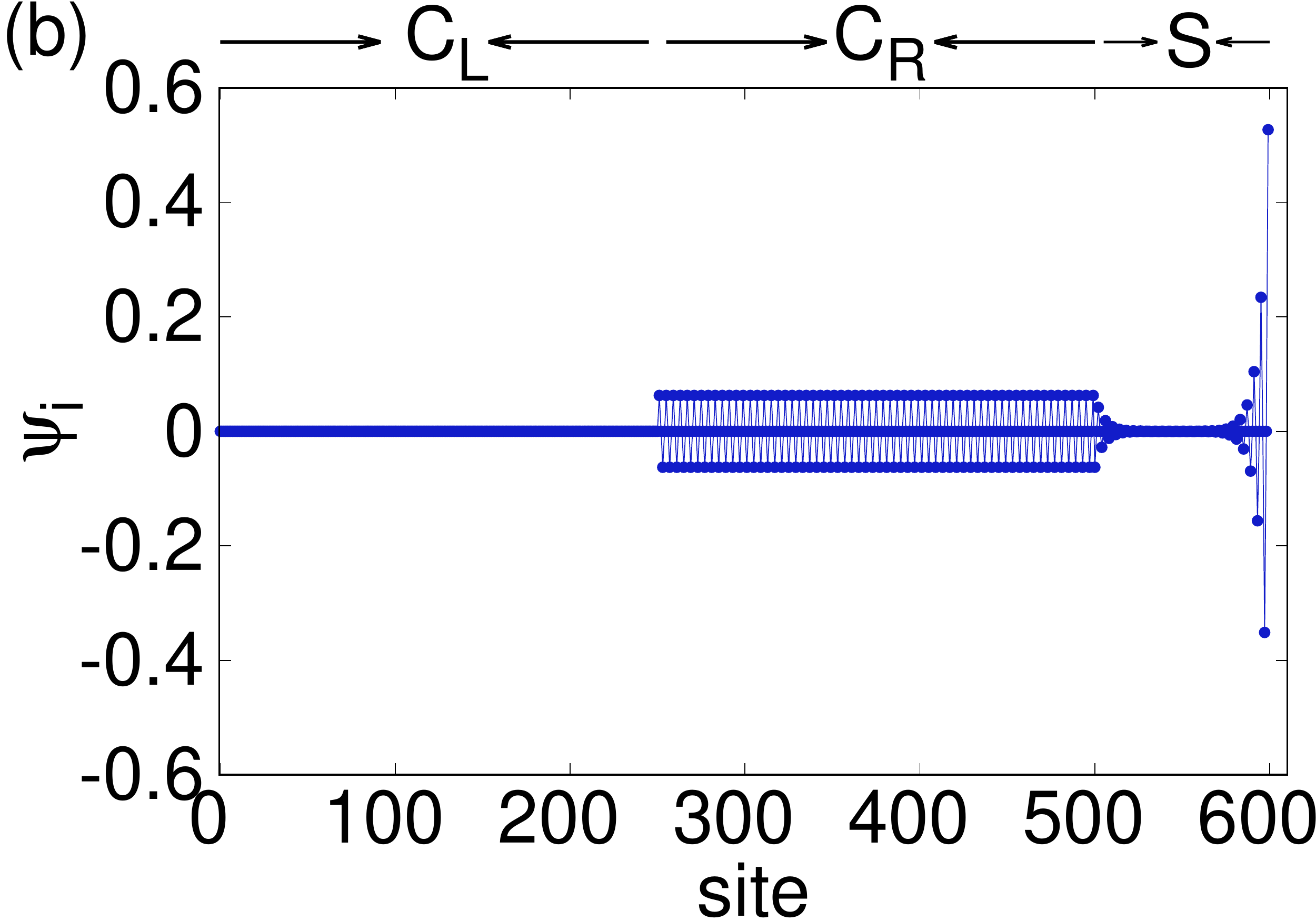}
\caption{The wavefunction for the two zero energy states with a simple
  tight binding chain with $N_c=500$ (site indices running from $1$ to
  $500$) as main chain and an open SSH chain with $N=100$ (site
  indices running from $501$ to $600$) as a side chain which is
  connected to one of the middle sites. The connecting hopping
  strengths are $v_1=1.0$, $v_2=0$ and other parameters are $v=1.0$,
  $t=1.0$ and $t'=1.5$.}
\label{wfxn_tbeo_1}
\end{figure}

The isolated SSH open chain with an even number of sites shows zero
modes only when $t'>t$, and so we look at the case $t=1.0, t'=1.5$.
The system now possesses two zero energy states (which are primarily
represented at the two edges).  \fref{wfxn_tbeo_1} shows the site
coefficients of the two zero energy eigenstates for the full
Hamiltonian containing the main chain and the side chain, with a
single connection $v_1$ between them.
We observe that both these states possess a finite probability amplitude one part of the
main chain and also at the free end of the SSH chain as shown in
\fref{wfxn_tbeo_1}. The zero amplitude of the eigenstate in one half
of the main chain confirms the full reflection in the system
corresponding to this energy as depicted in \fref{Ts_eo}.

\begin{figure}[h!]
\includegraphics[scale=0.168,angle=-90]{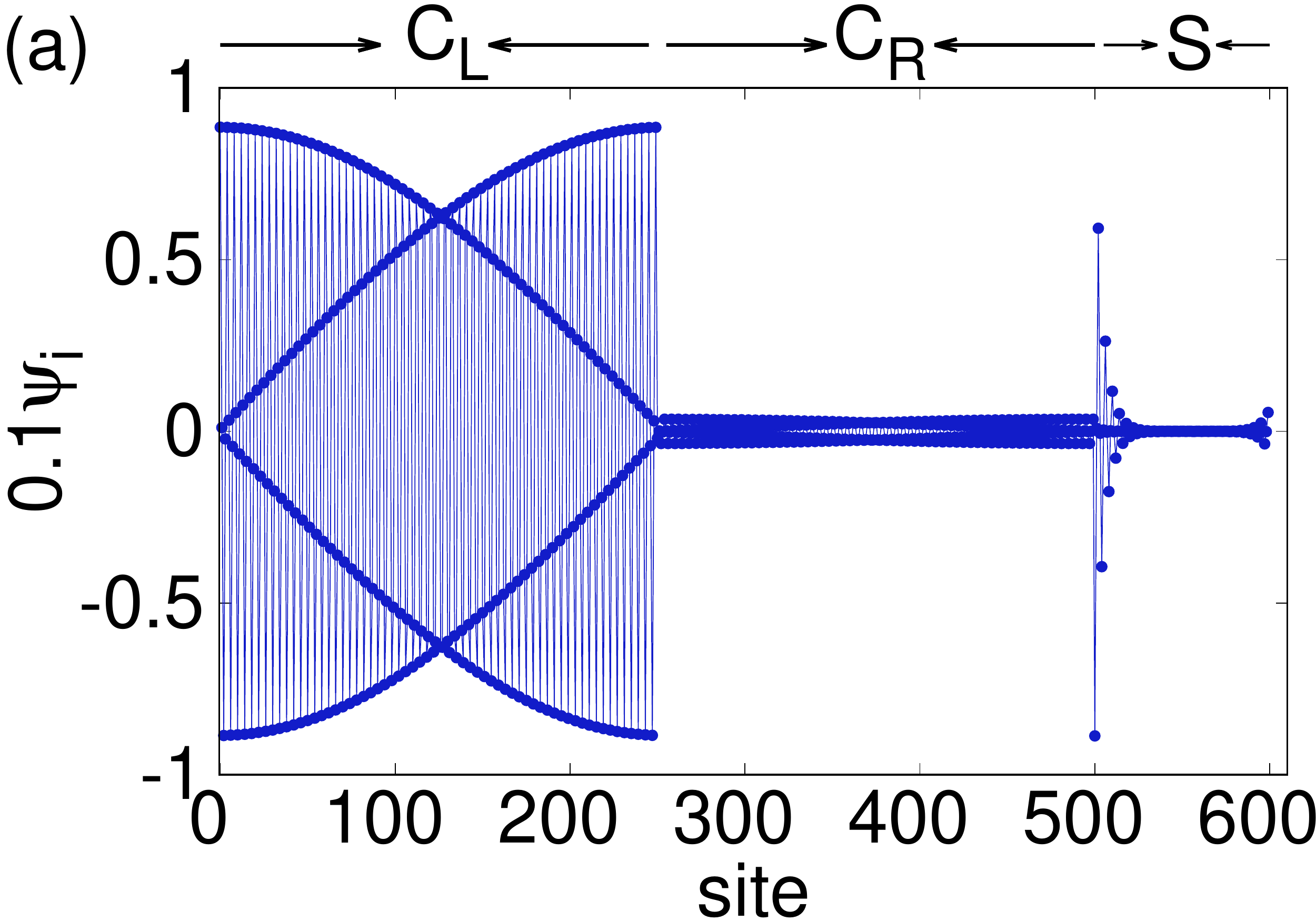}
\includegraphics[scale=0.168,angle=-90]{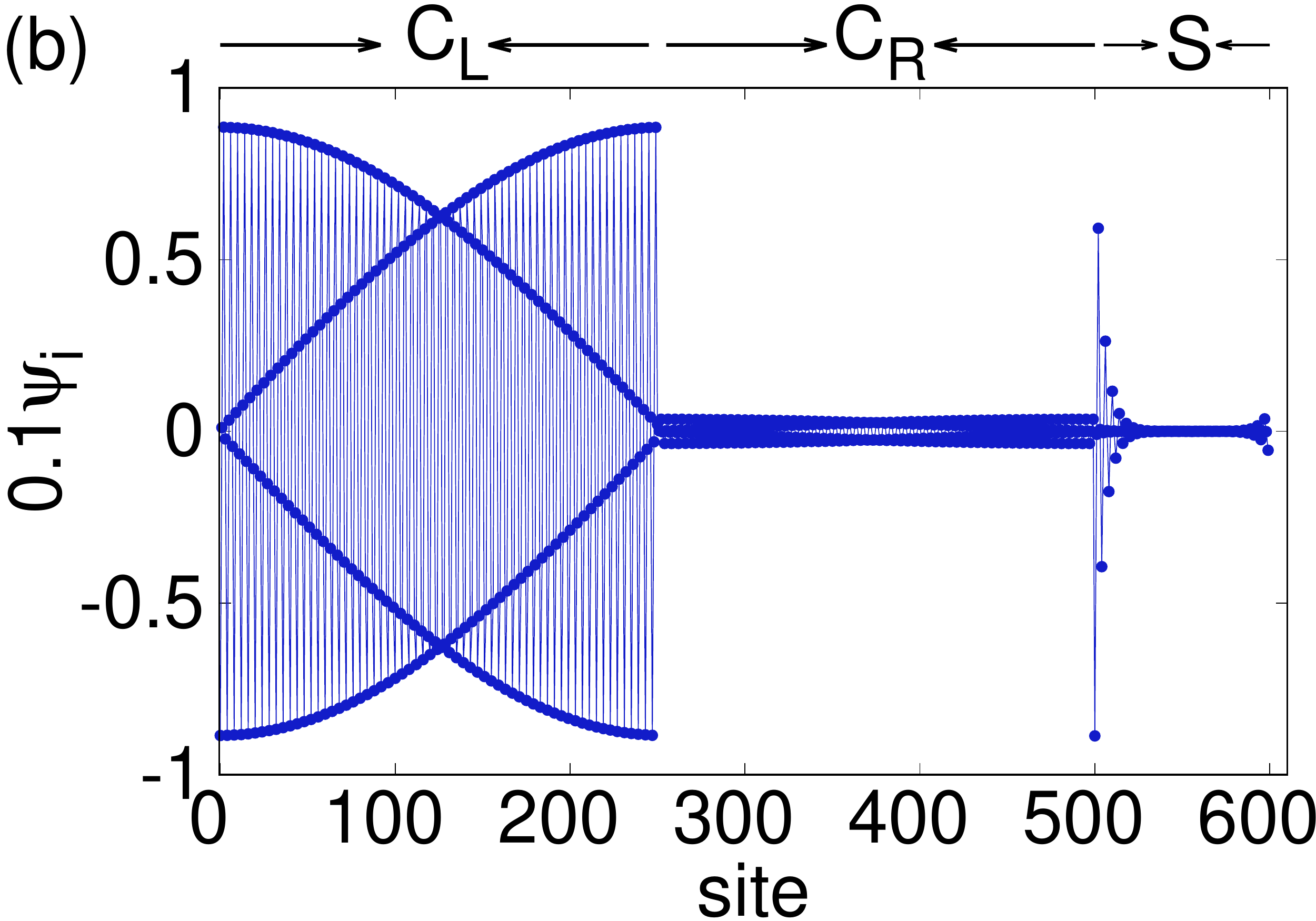}
\caption{The wavefunction for the two zero energy states for system
  consisting of a simple tight binding chain with $N_c=500$ (site
  indices running from $1$ to $500$) as main chain and an open SSH chain
  with $N=100$ (site indices running from $501$ to $600$) as side chain
  which is connected to one of the middle sites and an adjacent
  site. The connecting hopping strengths are $v_1=1.0$, $v_2=1.0$ and
  the other parameters are $v=1.0$, $t=1.0$ and $t'=1.5$.}
\label{wfxn_tbeo_2}
\end{figure}

The presence of the second connection $v_2$ leaves no free end for the
SSH chain.  Thus, the eigenstate for the full system is modified as
shown in \fref{wfxn_tbeo_2}.
The majority of the probability amplitude lies on the one half of the
main chain which results in zero transport in the system as depicted in
\fref{Td_eo}.


\section{Nearest neighbor tight binding open chain as side unit}
\label{lin_chain}
\begin{figure}[H] \centering
\includegraphics[scale=0.6]{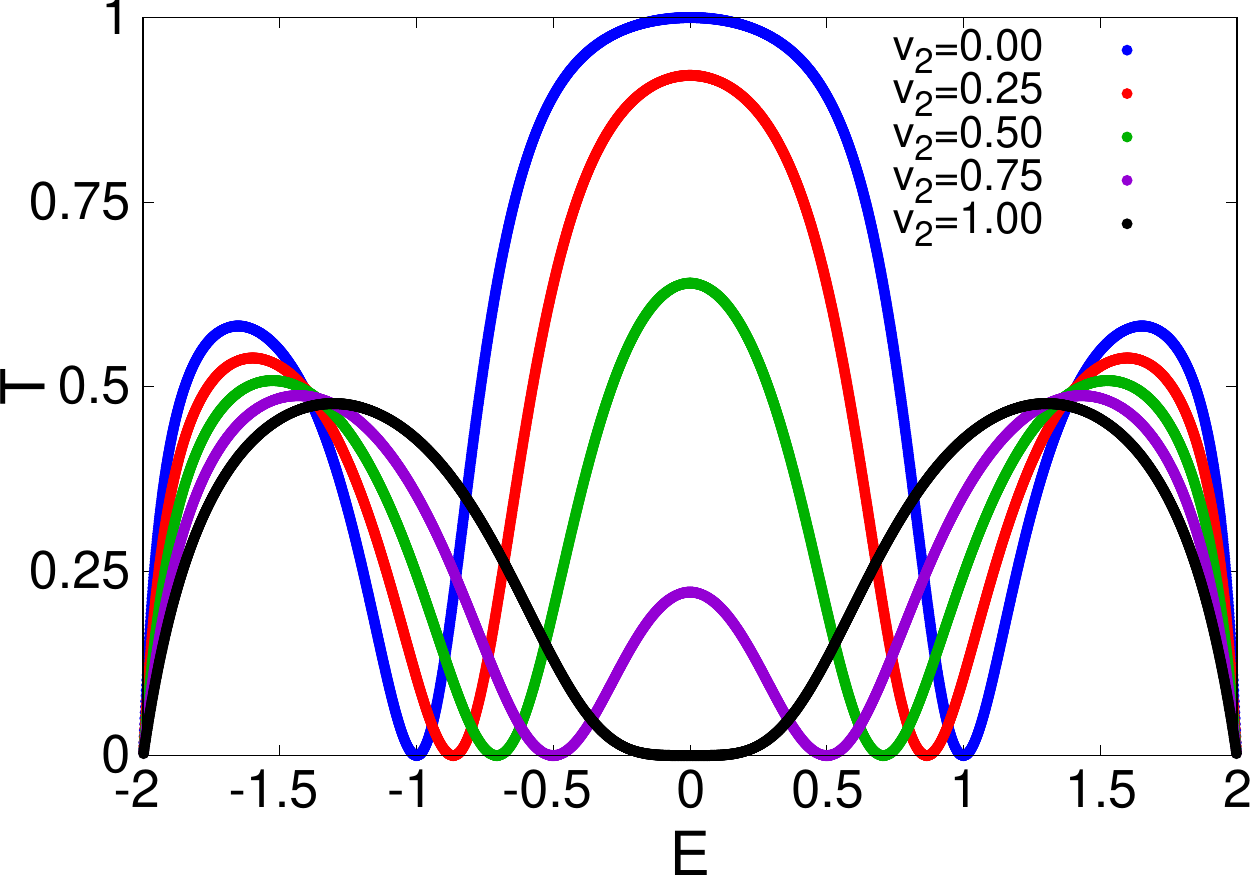}
\caption{Transmission coefficient as a function of incoming wave
  energy with various second connection strengths $v_2$ when the main
  chain is the simple tight binding chain and the side unit is a
  two-site chain. The other parameters are $v=1.0$, $v_1=1.0$,
  $t=1.0$.}
\label{Fano_tbo_2}
\end{figure}

We consider a nearest neighbour tight binding chain with hopping $t$
(i.e., $t=t'$) coupled to the main chain with two connections $v_1$
and $v_2$. %
\begin{figure}[h!]
\includegraphics[scale=0.6]{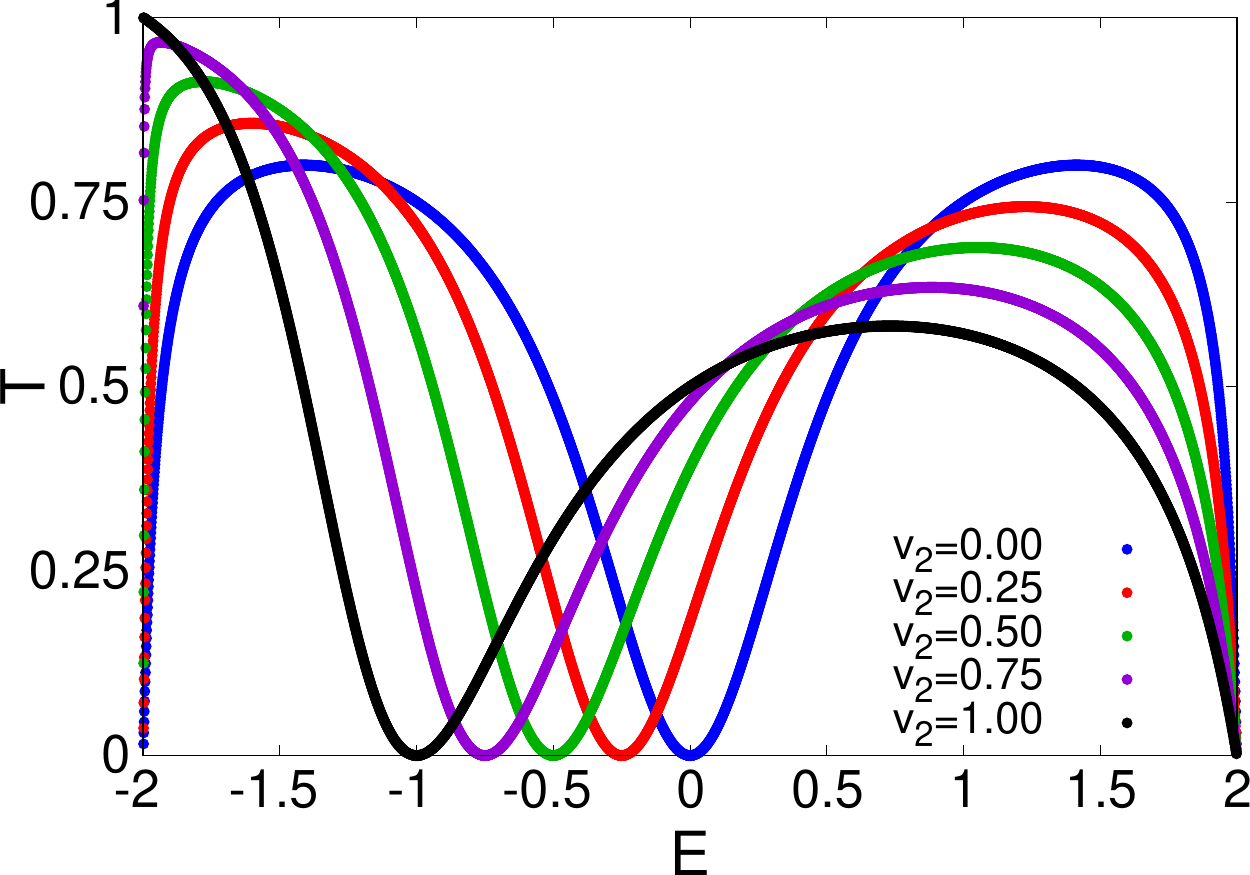}
\caption{Transmission coefficient as a function of the incoming wave
  energy with various second connection strengths $v_2$ when the main
  chain is the simple tight binding chain and the side unit is a
  single site. The parameters are $v=1.0$, $v_1=1.0$, $t=1.0$.}
\label{Fano_tbo_1}
\end{figure}
The transmission coefficient ($T$) as a function of
incoming wave energies ($E$) is shown in \fref{Fano_tbo_1} and
\fref{Fano_tbo_2} for different secondary connection strength $v_2$
when the sideunit possesses one ($N = 1$) and two ($N=2$) sites
respectively.

In the absence of $v_2$, the system shows a minimum transmission ($T=0$)
at $E=0$ when the side unit contains an odd number of sites as the
condition $\Gamma_N=0$ is satisfied. For even number of sites in the
side chain, the system shows a maximum transmission ($T=1$) at $E=0$
as the condition $\Gamma_{N-1}^{\{1\}}=0$ is satisfied.

In the presence of the second connection $v_2$, the perfect reflection
condition is given by:
\begin{equation}
v\Gamma_N+v_1v_2t^{N-1}=0,
\label{App_eq1}
\end{equation}
where $\Gamma_N=\displaystyle\prod_{k}(E-2t\cos(k))$ with
$k=\frac{j\pi}{N+1}$; $j=1,\hdots,N$ with $N$ being the side chain
length.
If a single defect site is connected to the main chain through two
connections $v_1$ and $v_2$ then the position of the minimum of
transmission is shifted by $-v_2$ ( \fref{Fano_tbo_1}) according to
\eref{App_eq1}. Again invoking \eqref{App_eq1} for $N=2$, as $v_2$ is increased, the minima at
$E=\pm 1$ move towards each other as shown in \fref{Fano_tbo_2}. Thus,
the maximum at $E=0$ for $v_2=0$ turns into a minimum for $v_2 = 1$.


\end{document}